\documentclass[prd,superscriptaddress,nofootinbib,twocolumn]{revtex4-2}

\usepackage{graphicx}
\usepackage{amsmath}
\usepackage{BOONDOX-uprscr}
\usepackage{amssymb}
\usepackage{amsfonts}
\usepackage{booktabs}
\usepackage{multirow}
\usepackage{url}
\usepackage{tabularx}
\usepackage{array}
\usepackage{siunitx}
\usepackage{xcolor}
\usepackage[colorlinks=true, allcolors=blue]{hyperref}
\usepackage{etoolbox}

\newcommand{\trm}{\textrm}

\newcommand{\Unu}{\ensuremath{\mathrm{UHE}\nu}}

\newcommand{\atUC}{\affiliation{Dept.~of Physics, Dept.~of Astronomy and Astrophysics, Enrico Fermi Institute, Kavli Institute for Cosmological Physics, University of Chicago, Chicago, IL 60637}}
\newcommand{\atKU}{\affiliation{Dept.~of Physics and Astronomy, University of Kansas, Lawrence, KS 66045}}
\newcommand{\atOSU}{\affiliation{Dept.~of Physics, Center for Cosmology and AstroParticle Physics, The Ohio State University, Columbus, OH 43210}}
\newcommand{\atUW}{\affiliation{Dept.~of Physics, Wisconsin IceCube Particle Astrophysics Center, University of Wisconsin-Madison, Madison,  WI 53706}}
\newcommand{\atNTU}{\affiliation{Dept.~of Physics, Graduate Institute of Astrophysics, Leung Center for Cosmology and Particle Astrophysics, National Taiwan University, Taipei, Taiwan}}
\newcommand{\atULB}{\affiliation{Universit\'{e} Libre de Bruxelles, Science Faculty CP230, B-1050 Brussels, Belgium}}
\newcommand{\atUMD}{\affiliation{Dept.~of Physics, University of Maryland, College Park, MD 20742}}
\newcommand{\atPSUigc}{\affiliation{Center for Multi-Messenger Astrophysics, Institute for Gravitation and the Cosmos, Pennsylvania State University, University Park, PA 16802}}
\newcommand{\atPSUphys}{\affiliation{Dept.~of Physics, Pennsylvania State University, University Park, PA 16802}}
\newcommand{\atPSUast}{\affiliation{Dept.~of Astronomy and Astrophysics, Pennsylvania State University, University Park, PA 16802}}
\newcommand{\atVUB}{\affiliation{Vrije Universiteit Brussel, Brussels, Belgium}}
\newcommand{\atUNL}{\affiliation{Dept.~of Physics and Astronomy, University of Nebraska, Lincoln, Nebraska 68588}}
\newcommand{\atWhittier}{\affiliation{Dept.~Physics and Astronomy, Whittier College, Whittier, CA 90602}}
\newcommand{\atNUU}{\affiliation{Dept.~of Energy Engineering, National United University, Miaoli, Taiwan}}
\newcommand{\atChiba}{\affiliation{International Center for Hadron Astrophysics, Chiba University, Chiba 263-8522 Japan}}
\newcommand{\atNPU}{\affiliation{Dept.~of Applied Physics, National Pingtung University, Pingtung City, Pingtung County 900393, Taiwan}}
\newcommand{\atUCL}{\affiliation{Dept.~of Physics and Astronomy, University College London, London, United Kingdom}}
\newcommand{\atUD}{\affiliation{Dept.~of Physics, University of Delaware, Newark, DE 19716}}
\newcommand{\atUNAF}{\affiliation{Unaffiliated, Madison, WI 53706}}
\newcommand{\atNDL}{\affiliation{National Nano Device Laboratories, Hsinchu 300, Taiwan}}
\newcommand{\atUtah}{\affiliation{Dept.~of Physics and Astronomy, University of Utah, Salt Lake City, UT 84112}}
\newcommand{\atDESY}{\affiliation{Deutsches Elektronen-Synchrotron DESY, Platanenallee 6, 15738 Zeuthen, Germany}}

\begin{document}


\title{Analysis Methodology of the Search for Ultrahigh Energy Neutrinos\\in 10.6 Years of Askaryan Radio Array Data}


 \author{N.~Alden}\atUC
 \author{S.~Ali}\atKU
 \author{P.~Allison}\atOSU
 \author{D.Z.~Besson}\atKU
 \author{A.~Bishop}\email{abigail.bishop@icecube.wisc.edu}\atUW
 \author{P.~Chen}\atNTU
 \author{Y.C.~Chen}\atNTU
 \author{Y.-C.~Chen}\atNTU
 \author{S.~Chiche}\atULB
 \author{B.A.~Clark}\email{baclark@umd.edu}\atUMD
 \author{A.~Connolly}\atOSU
 \author{K.~Couberly}\atKU
 \author{A.~Cummings}\atPSUigc\atPSUphys\atPSUast
 \author{P.~Dasgupta}\email{dasgupta.80@osu.edu}\atOSU
 \author{K.D.~de~Vries}\atVUB
 \author{C.~Deaconu}\atUC
 \author{M.A.~DuVernois}\atUW
 \author{J.~Flaherty}\atOSU
 \author{P.~Giri}\email{pgiri4@huskers.unl.edu}\atUNL
 \author{J.~Hanson}\atWhittier
 \author{K.D.~Hoffman}\atUMD
 \author{M.-H.~Huang}\atNTU\atNUU
 \author{K.~Hughes}\atOSU
 \author{A.~Ishihara}\atChiba
 \author{A.~Karle}\atUW
 \author{J.L.~Kelley}\atUW
 \author{M.-C.~Kim}\atChiba
 \author{I.~Kravchenko}\atUNL
 \author{R.~Krebs}\atPSUigc\atPSUphys
 \author{C.Y.~Kuo}\atNTU
 \author{U.A.~Latif}\atVUB
 \author{C.H.~Liu}\atUNL
 \author{T.C.~Liu}\atNTU\atNPU
 \author{W.~Luszczak}\atOSU
 \author{A.~Machtay}\email{machtay.1@osu.edu}\atOSU
 \author{M.S.~Muzio}\email{muzio@wisc.edu}\atUW
 \author{J.~Nam}\atNTU
 \author{R.J.~Nichol}\atUCL
 \author{A.~Novikov}\atUD
 \author{A.~Nozdrina}\atOSU
 \author{E.~Oberla}\atUC
 \author{C.W.~Pai}\atNTU
 \author{Y.~Pan}\atUD
 \author{C.~Pfendner}\atUNAF
 \author{N.~Punsuebsay}\atUD
 \author{J.~Roth}\atUD
 \author{A.~Salcedo-Gomez}\email{salcedogomez.1@osu.edu}\atOSU
 \author{D.~Seckel}\atUD
 \author{M.F.H.~Seikh}\email{fulhossain@ku.edu}\atKU
 \author{Y.-S.~Shiao}\atNTU\atNDL
 \author{S.C.~Su}\atNTU
 \author{S.~Toscano}\atULB
 \author{J.~Torres}\atOSU\atUtah
 \author{N.~van~Eijndhoven}\atVUB
 \author{A.~Vieregg}\atUC
 \author{M.~Vilarino~Fostier}\atULB
 \author{M.-Z.~Wang}\atNTU
 \author{S.-H.~Wang}\atNTU
 \author{P.~Windischhofer}\atUC\atDESY
 \author{S.A.~Wissel}\atPSUigc\atPSUphys\atPSUast
 \author{S.~Yoshida}\atChiba
\collaboration{ARA Collaboration}\noaffiliation

\date{\today}

\begin{abstract}
The Askaryan Radio Array (ARA) is an in-ice radio observatory at the South Pole designed to detect ultrahigh energy (UHE) neutrinos through Askaryan emission produced by interactions of neutrinos and their secondary particles. In this paper, we present the detailed methodology of the first array-wide search for a diffuse UHE neutrino flux using data from all five ARA stations. The dataset considered represents $10.6$\,years of livetime and is three times larger than previous ARA neutrino searches. This search achieves a $28\%$~analysis efficiency for a neutrino flux corresponding to the current experimental upper limits. No neutrino candidates are observed on an expected background of $0.13\pm0.01\text{ (stat.)}^{+0.05}_{-0.06}\text{ (syst.)}$~events. We set a new upper limit on the diffuse UHE neutrino flux, which is the most stringent from $50$\,EeV to $1000$\,EeV. We also discuss lessons learned for future searches on large radio arrays.
\end{abstract}

\maketitle

\onecolumngrid          
\clearpage             
\twocolumngrid          

\section{Introduction}\label{sec:intro}

The origin of ultrahigh energy cosmic rays (UHECRs; $E_\mathrm{CR}\gtrsim 1$\,EeV) remains an unresolved question in astroparticle physics.
Spectrum, composition, and anisotropy measurements of UHECRs indicate they are predominantly extragalactic and composed of heavy nuclei at the highest energies~\cite{Coleman:2022abf,Roth:2025hqe,Kido:2025mey}. 
However, due to deflections in Galactic and extragalactic magnetic fields~\cite{Unger:2023lob,Bister:2024ocm,PierreAuger:2024hlp},
source identification remains a challenge.

Ultrahigh energy neutrinos (\Unu{}; $E_{\nu}\gtrsim 100$\,PeV) provide a complementary probe of UHECR sources.
Since they are electrically neutral and weakly interacting, neutrinos 
do not suffer from magnetic deflections or significant attenuation in propagation, making them ideal particle messengers.
Moreover, UHE neutrinos are produced in UHECR interactions, making them a direct probe of UHECR sources.

The diffuse neutrino flux is expected to consist of astrophysical neutrinos, produced through interactions inside their sources, and cosmogenic neutrinos, produced during extragalactic propagation.
Astrophysical neutrinos are generated both hadronically and photohadronically within or near their sources.
Cosmogenic neutrinos are produced through photopion production interactions of UHECRs with the cosmic microwave background~(CMB) and extragalactic background light~(EBL). In particular, the proton flux above about $10^{19.7}$\,eV is suppressed by these interactions with the CMB via the Greisen--Zatsepin--Kuzmin~(GZK) effect~\cite{Greisen1966,Zatsepin1966}.
Therefore, the total diffuse \Unu{} flux provides a probe of both the chemical composition of UHECRs and the conditions inside their sources. 

The KM3NeT experiment recently observed the first \Unu{} candidate, with an energy of $220^{+570}_{-110}$\,PeV~\cite{KM3NeT:2025npi}.
At higher energies, the strongest constraints on the diffuse flux of \Unu{} have been set by IceCube~\cite{IceCubeCollaborationSS:2025jbi} and ANITA~\cite{ANITA:2019wyx} at the level of $E_{\nu}^{2 }\Phi_{\nu} \lesssim 10^{-8}$\,GeV/cm$^{2}$/s/sr near $E_{\nu}=1$\,EeV.
Greater sensitivity to the \Unu{} flux requires monitoring tens to hundreds of cubic kilometers of target material over extended periods.
In-ice radio arrays provide a scalable approach to achieve this by detecting Askaryan emission~\cite{Askaryan:1961pfb,ANITA:2006nif} from neutrino-induced particle cascades. Such detectors work by deploying clusters of antennas called ``stations" in the ice, many of which compose a detector array.
Antarctic ice is an ideal medium for such arrays due to its long $\mathcal{O}(1)$~km attenuation lengths at radio frequencies~\cite{Barwick:2005zz}.
For this reason, Antarctica has been host to several experiments designed to leverage this technique, including the Radio Ice Cherenkov Experiment (RICE)~\cite{RICE:2001ayk}, the Antarctic Ross Ice-Shelf Antenna Neutrino Array (ARIANNA)~\cite{Barwick:2016mxm}, and the Askaryan Radio Array (ARA)~\cite{ARA:2011testbed}. Other Askaryan detector designs include long-duration balloons flown over the Antarctic sheet, such as ANITA and the recently-flown PUEO~\cite{PUEO:2020bnn,PUEO:2026owv}.

In this work and the accompanying Letter~\cite{prl}, we present the first array-wide diffuse \Unu{} search performed with ARA.
This analysis combines data from all five ARA stations, collected over $10.6$\,years, yielding a dataset more than three times larger than those used in previous ARA neutrino searches~\cite{ARA:2019wcf,ARA:2011testbed,ARA:2022rwq}. 
No neutrino candidates are observed, and we set a $90\%$~confidence level upper limit on the diffuse \Unu{} flux, which is the strongest from $50$\,EeV to $1000$\,EeV.
This search also establishes an array-wide analysis strategy that can be adopted by future large-scale radio observatories.
In Section~\ref{sec:ARA} we describe the ARA detector and the dataset used for this search.
We present the neutrino and detector simulations used to determine the sensitivity of this analysis in Section~\ref{sec:simulations}.
In Section~\ref{sec:data_proc}, we detail the event processing and reconstruction. These are utilized in the final event selection and expected background estimation described in Section~\ref{sec:event_selection}. In Section~\ref{sec:uncertainties}, we discuss the statistical and systematic uncertainties on this result.
Finally, we present our results in Section~\ref{sec:results}, discuss their implications in Section~\ref{sec:discussion}, and give our conclusions in Section~\ref{sec:conclusion}.

\begin{figure}[h!]
    \centering
    \includegraphics[width=\linewidth]{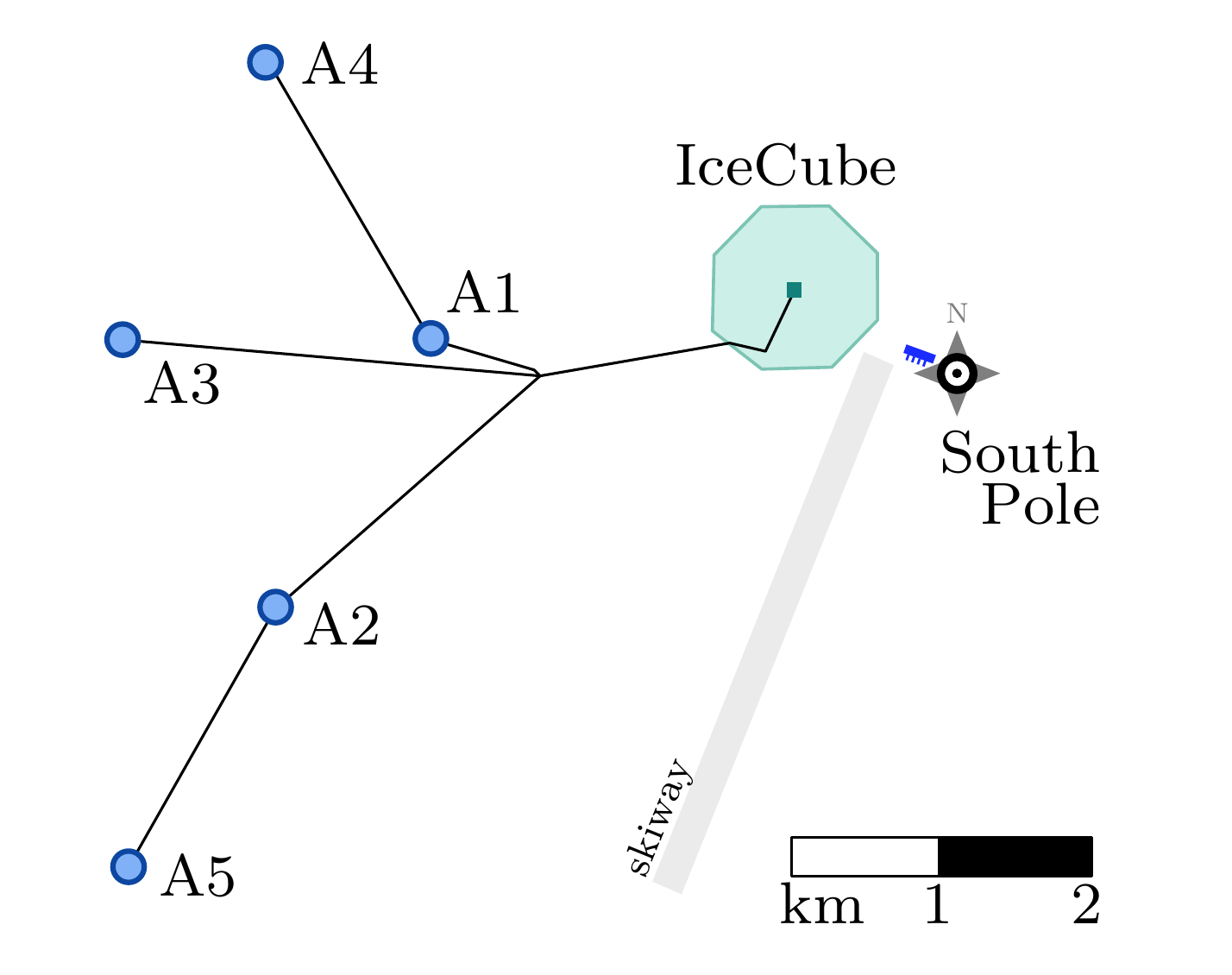}
    \caption{
        Map of the five ARA stations relative to the IceCube Neutrino Observatory and South Pole. The A5 site is also the site of the phased array detector. The compass marker denotes grid north.
        \label{fig:ARA_map}
    }
\end{figure}

\section{ARA}\label{sec:ARA}

ARA stations, each consisting of a compact cluster of in-ice radio antennas, are distributed on a roughly $2$~km-spaced hexagonal grid located near the South Pole and the IceCube Neutrino Observatory, as shown in Fig.~\ref{fig:ARA_map}. The array consists of five autonomous stations, labeled A1--A5, embedded in the Antarctic glacier.
\Unu{} interacting in the Antarctic ice initiate relativistic particle cascades. 
As the cascade develops, interactions with the surrounding medium produce a net negative charge excess, which generates a coherent, nanosecond-scale radio pulse along the Cherenkov cone.
This pulse, known as Askaryan radiation~\cite{Askaryan:1961pfb}, was first observed using an electron beam at SLAC~\cite{ANITA:2006nif} and recently observed in-situ by ARA through cosmic ray-induced showers penetrating the ice~\cite{ARA:2025qit}.

\begin{figure}[h!]
    \centering
    \includegraphics[width=\linewidth]{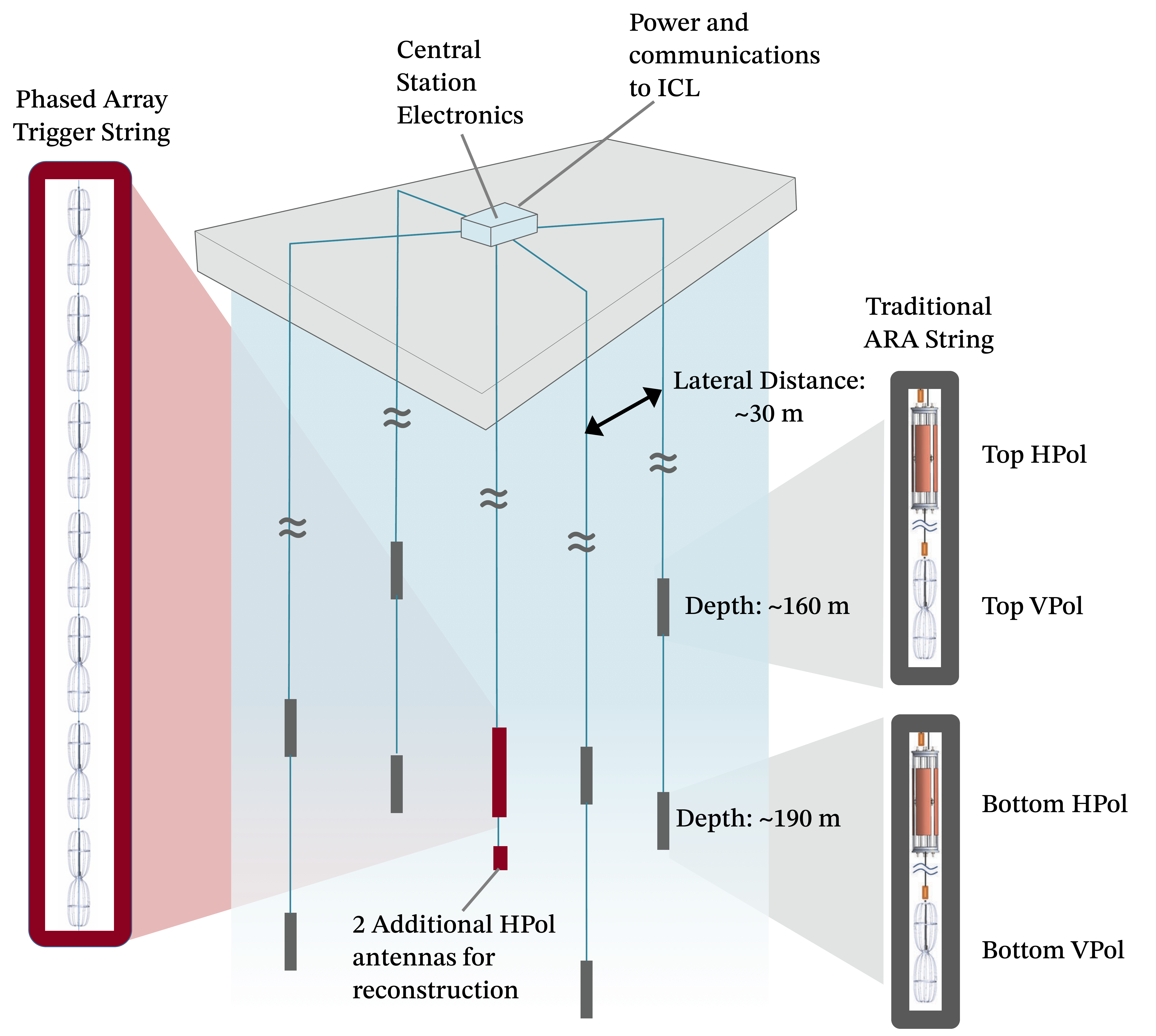}
    \caption{
        ARA's station layout, shown for station A5. Each ARA station has four holes drilled into the ice and instrumented with radio antennas (shown in grey), extending to a maximum depth of $90$\,m at A1 and approximately $200$\,m at A2--A5. A5 additionally has the phased array (PA) detector string (shown in red) deployed in the center of the four outer strings. A1--A3 have an installed lateral distance baseline of $14$\,m, while A4 and A5 have a $30$\,m baseline.
        \label{fig:ARA_station}
    }         
\end{figure}

\subsection{Detector}

Each ARA station operates independently and records radio-frequency (RF) signals.
A prototype station, known as the ``Testbed", was deployed during the 2010--2011 season to characterize the radio environment and validate the detector concept~\cite{ARA:2011testbed}.
The first station, A1, was installed during the 2011--2012 season with antennas deployed up to $90$\,m depth.
Stations A2 and A3 followed during the 2012--2013 season with antennas deployed up to $190$\,m depth.
Two additional deep stations, A4 and A5, were installed during the 2017--2018 season.

A standard ARA station consists of four boreholes arranged in an approximately square footprint, as shown in Fig.~\ref{fig:ARA_station}, with lateral separations of approximately $14$\,m for A1--A3 and $30$\,m for A4 and A5.
Each borehole hosts four receiving antennas, two sensitive to ``vertically'' polarized signals (VPols) and two sensitive to ``horizontally'' polarized signals (HPols).
Specifically, VPols are mostly sensitive to the $\hat{\theta}$ component of the incident electric field, which lies in the vertical plane containing the signal arrival direction.
HPols are mostly sensitive to the orthogonal $\hat{\phi}$ component, which lies in the horizontal plane.\footnote{These directions are defined in a local spherical coordinate system $(r,\theta,\phi)$, with $\hat{r}$ pointing from the antenna toward the source, opposite to the local propagation vector $\mathbf{k}$ of the received signal.}
The VPol antennas are wire-frame bicones, while the HPol antennas are quad-slot cylinders.
Both antenna types are designed to fit within the narrow $15$\,cm borehole diameter while maintaining broadband sensitivity over the frequency range relevant for Askaryan emission, approximately $150-850$\,MHz.
Each station also includes up to two calibration strings, instrumented with one VPol and one HPol transmitting antenna.
These pulsers are used for timing and geometry calibration, and for long-term monitoring of detector stability. 

\begin{figure*}
    \centering
    \includegraphics[width=\linewidth]{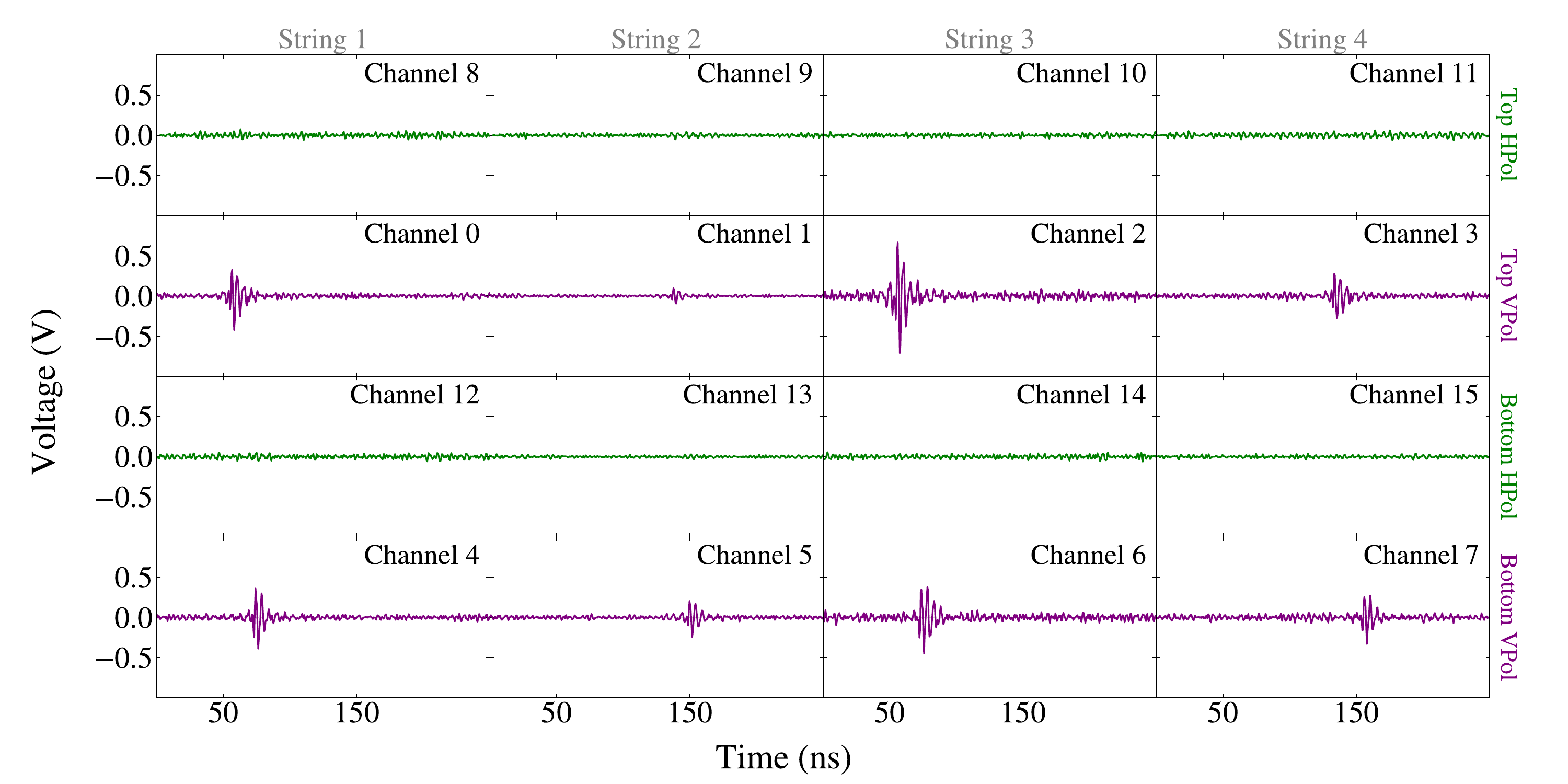}
    \caption{Voltage traces for a calibration pulser event. The strings of the detector are labeled, and each channel is colored by its polarization with VPols (purple) and HPols (green). The rows correspond to depth in the detector, with the top row being the ``top HPol" antennas, the second row being the ``top VPol" antennas, etc.
    This corresponds to the detector layout presented in Fig.~\ref{fig:ARA_station}.}
    \label{fig:example_cp_event}
\end{figure*}

When an antenna receives a signal, it is amplified and then transported to the surface via fiber links.
After additional filtering and amplification by the surface electronics, the signal is split into trigger and digitization paths.
In the standard A1--A5 stations, the signal in each channel is passed through a tunnel-diode power integrator, which generates a single-channel trigger when the integrated power exceeds a threshold.
A station-wide trigger is issued when at least three antennas of the same polarization trigger within a coincidence window, typically $\sim170$\,ns, which corresponds to the light travel time across the station.
The single-channel thresholds are dynamically adjusted to maintain a trigger rate near $5$\,Hz per station.
Stations also periodically record software-forced triggers at approximately $1$\,Hz.
Calibration pulses are also recorded at $1$\,Hz, yielding a total trigger rate of about $7$\,Hz.
Upon triggering, sampling is halted, and the analog buffers are read out, recording a waveform window of approximately $500$\,ns.
The data acquisition system subsequently undergoes a readout deadtime of approximately $1$\,ms, corresponding to a sub-percent livetime loss at nominal trigger rates.
The set of sixteen waveforms, along with associated housekeeping data, becomes an ``event."
An example calibration pulser event is displayed in Fig.~\ref{fig:example_cp_event}.

\begin{figure*}
    \centering\includegraphics[width=\textwidth]{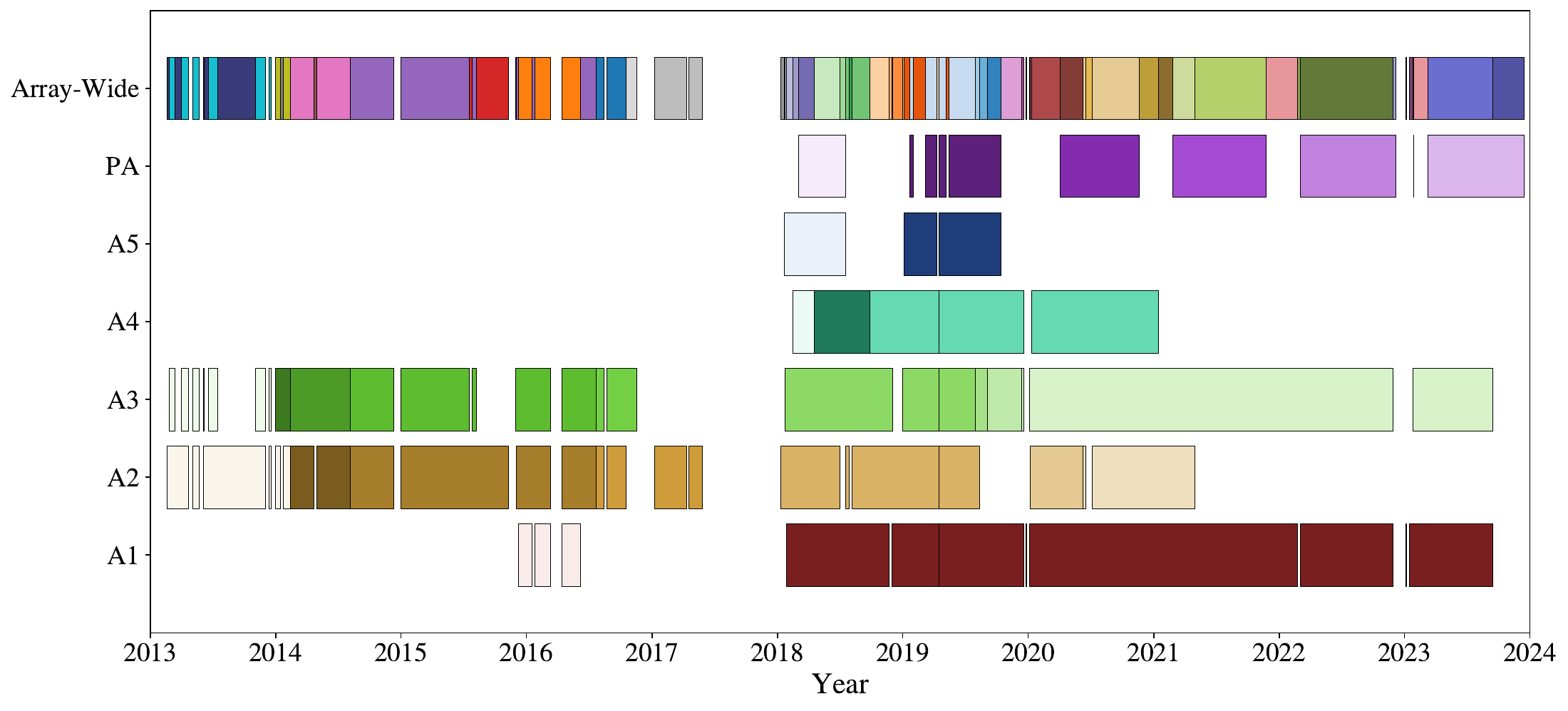}
    \caption{ARA detector configurations from January 2013 to December 2023. The lower rows show station-level configurations for A1--A5 and PA, while the top row shows the resulting array-wide configurations. Repeated colors in a row indicate the same configuration, while gaps denote periods when the station was not taking data or the livetime was excluded from the analysis.}
    \label{fig:arraywide_configs}
\end{figure*}

A5 also hosts a phased array (PA) trigger system consisting of a central string of seven VPol antennas and two HPol antennas with approximately $1-2$\,m spacing, as shown in  Fig.~\ref{fig:ARA_station}.
This design enables coherent beamforming of the signals to achieve a lower threshold than the standard coincidence trigger.
The PA trigger is determined on a sum of time-delayed voltage traces from the VPol antennas~\cite{ARA:2018phased}.
These time delays correspond to the relative arrival times of a plane wave from predefined directions~\cite{Vieregg:2015baa}.
Signals aligned with a given direction add coherently, scaling with the number of channels $N$, while uncorrelated noise adds incoherently, scaling with $\sqrt{N}$.
This amplifies the signal-to-noise ratio (SNR) of coherent signals.
The PA system sums signals along $15$~directions, with thresholds dynamically adjusted to maintain a rate of approximately $0.75$\,Hz per direction and a global trigger rate of $11$\,Hz~\cite{ARA:2018phased}.
The PA reaches $50\%$~trigger efficiency at a single-antenna voltage SNR of approximately 2.0, compared to 3.7 for the coincidence trigger~\cite{ARA:2022rwq}.
The PA trigger and readout operate independently from the co-located standard A5 detector, and therefore the PA is treated as a separate instrument in this analysis. The A5 readout, on the other hand, was triggered both by its standard coincidence trigger and by external triggers from the PA detector. 
During the 2018--2019 season, one A5 VPol channel was added to the PA data acquisition system. In the 2019--2020 season, the A5 system was decommissioned and merged with the PA system. This improved reconstruction capability by adding six of the VPol channels from A5 to the PA system, which was the maximum number of available spare readout channels in the PA system.
These ``merged" channels did not participate in the triggering and were only used for readout.

\subsection{Dataset}

\subsubsection{Livetime}

The dataset considered in this search spans January 2013 through December 2023, and includes data from all five ARA stations, including the PA detector.
Before performing data quality selections (described in Section~\ref{sec:event_selection}), the dataset comprises approximately $28$\,station-years.
After these selections, $25.9$\,station-years remain, corresponding to $10.6$\,years of array-wide livetime (i.e.\ the total livetime during which at least a portion of the array was taking data).

To reduce biases during analysis development, we employ a data prescaling blinding strategy~\cite{Klein:2005di}.
A randomly selected $10\%$ sample was used to characterize background populations and develop the event selection criteria, while the remaining $90\%$ was kept blinded until the selection and expected background estimates were finalized.

\subsubsection{Detector Configurations}

The detector's operating parameters change as a function of time. We parametrize these as ``detector configurations'' to account for time-dependent changes in each station's response due to variations in noise conditions, trigger thresholds, and other data acquisition settings relevant to detector modeling and data analysis.
The dataset of each station is divided into time periods, which we call ``station-level configurations", over which these properties are approximately stable.
We further define ``array-wide configurations" by the combination of active station-level configurations at a given time.
The evolution of station-level and array-wide configurations over time is shown in Fig.~\ref{fig:arraywide_configs}.
The array-wide configurations are used for detector simulations and exposure calculations.
For array-wide configurations in which both A5 and PA are active, we analyze only PA data to avoid treating the overlapping datasets as independent, and only simulate the PA.

\section{Simulations}\label{sec:simulations}

A two-stage Monte Carlo simulation and array-wide aggregation framework is used to evaluate ARA's \Unu{} sensitivity and the signal efficiency of the event selection (described in Section~\ref{sec:event_selection}).
First, primary neutrinos and their secondary particles are propagated through Earth using \texttt{NuLeptonSim}~\cite{Cummings:2023iuw} to form libraries of particle interactions.
Second, these interaction libraries are passed to \texttt{AraSim}~\cite{ARA:2026wjc} to simulate each station's response and triggering.
This results in a dataset of interaction vertices that triggered each station. Finally, these station-level datasets are aggregated array-wide by identifying triggered interactions originating from the same primary neutrino.
In this Section, we briefly expand on the description of these simulation stages; a detailed description can be found in~\cite{ARA:2026wjc}.

Neutrinos are generated by \texttt{NuLeptonSim} at discrete energies, in half-decade steps from $10^{16}$\,eV to $10^{21}$\,eV, and as an $E^{-1}$ spectrum, which is later re-weighted to various spectra of interest.
The fiducial volume for these simulations is defined by a cylinder centered on A2 (roughly the center of the array).
The cylinder has a $15$\,km radius and $3$\,km height, except for simulations at $E_\nu \leq 10^{17}$\,eV, which use smaller fiducial volumes (see~\cite{ARA:2026wjc} for details).
Primary neutrinos are generated along isotropically distributed chords that intersect uniformly distributed points in the fiducial volume.
Neutrinos are propagated through the Earth with interaction lengths determined using the neutrino-nucleon cross section of Connolly et al.~\cite{Connolly:2011vc} and a layer-averaged density profile derived from the Preliminary Reference Earth Model (PREM)~\cite{Dziewonski:1981xy}.

In addition to neutrinos, \texttt{NuLeptonSim} propagates secondary muons and taus resulting from primary neutrino interactions.
Muons and taus may generate additional hadronic and electromagnetic cascades through stochastic energy losses via bremsstrahlung, pair-production, and photonuclear processes~\cite{Cummings:2023iuw}.
Tau decays may also produce further detectable showers.
Consequently, a single primary neutrino can yield multiple interaction vertices within the fiducial volume that contribute to the number of array-wide triggers.
This effect becomes increasingly important at the highest energies~\cite{ARA:2026wjc}, where it is responsible for some $20-40\%$ of expected observed vertices.

The resulting interaction libraries are then passed to \texttt{AraSim} for station-level simulation of the signal generation and propagation to each antenna, as well as the detector response and trigger.
\texttt{AraSim} packages simulated events in the same format as real data, so the analysis pipeline described in Section~\ref{sec:data_proc} can be executed on both file types interchangeably.
Several updates, detailed in~\cite{ARA:2026wjc}, were made to \texttt{AraSim} to improve simulation fidelity as part of this analysis. 
We note that in stations A1--3, steeply upgoing signals have previously been observed to inefficiently trigger the detector~\cite{Allison:2019rgg}. 
The exact mechanism for this effect is not known and, therefore, is not captured by the trigger simulation.
For this reason, simulated neutrinos generating signals arriving at the antenna in this regime are excluded from the acceptance calculation for A1--3.

\begin{figure}
        \centering
        \includegraphics[width=\linewidth]{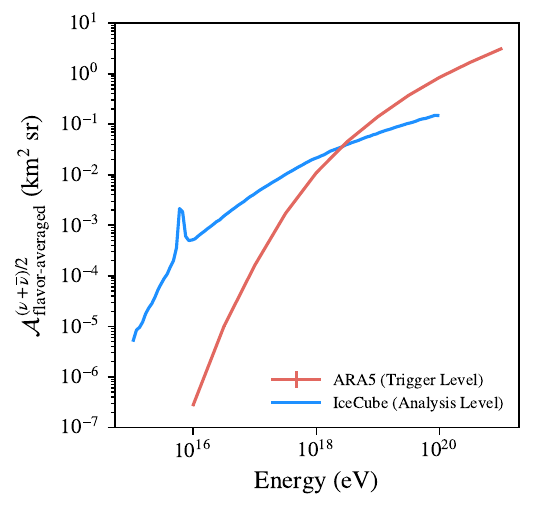}
        \caption{
            \label{fig:veffperstation}
             The trigger-level acceptance averaged over all six neutrino types and averaged over the livetime of the full array (red) compared to the analysis-level IceCube acceptance~\cite{IceCubeCollaborationSS:2025jbi} (blue). 
        }
    \end{figure}

Station-level simulations are produced by \texttt{AraSim} for each configuration, which are then aggregated to form array-wide simulation sets (see Fig.~\ref{fig:arraywide_configs}).
The resulting acceptance of the full array is shown in Fig.~\ref{fig:veffperstation}.

\section{Data Processing}\label{sec:data_proc}

The objective of the neutrino search is to apply a series of event selections (``cuts") to remove backgrounds and preserve neutrinos.
The most abundant background is thermal noise from the blackbody radiation of the environment and noise introduced by the system electronics, including the amplifiers.
Other, non-thermal-like backgrounds arise from humans (``anthropogenics")
and from physics, such as cosmic rays interacting in the ice and atmosphere.
The expected signal from a diffuse flux of neutrinos should be an impulsive, nanosecond-scale signal that originates within the ice and is observed in multiple antennas, while being time-isolated from other such signals.
Thermal and thermal-like events are expected to be non-impulsive, uncorrelated across channels, and weaker in overall amplitude.

Rather than analyzing raw data traces, higher-level analysis variables are calculated to characterize the event, such as its reconstructed arrival direction, signal strength, and impulsivity.
We describe the variable construction in this Section; the variables are then used in the event selection described in Section~\ref{sec:event_selection}.

\subsection{Event Pre-Processing}\label{sec:preprocessing}

We pre-process events using \texttt{AraProc}\footnote{\url{https://github.com/ara-software/AraProc}}, a dedicated framework for ARA data handling.
A station-dependent timing and voltage calibration is applied to each waveform~\cite{ARA:2023rtx,Meures:2014rfw,ARA:2021wmr}.
The calibrated waveforms are then interpolated onto a common 
time base to enable consistent channel-to-channel cross-correlations and frequency-domain processing.
Next, narrowband ``continuous wave" (CW) contamination from anthropogenic transmitters, such as NOAA weather radiosondes, is suppressed using a ``sine-subtraction" algorithm originally developed for ANITA~\cite{ANITA:2018impulsivity,ARA:2025scq}.
Finally, the frequency-dependent phase response of the amplifier and filter chain is deconvolved, and a $50-875$\,MHz fifth-order Butterworth bandpass filter is applied to remove out-of-band power.
The correction of the phase response reduces the dispersion introduced by the amplifiers and filters, rendering any contained signals more impulsive.

\subsection{Calculation of Analysis Variables}\label{sec:L2}

Following waveform pre-processing, we calculate event-level observables used in the subsequent event selection.
These observables serve two broad purposes: identifying analysis-quality events and distinguishing neutrino-like signals from background populations.
We first describe variables that characterize signal strength and impulsivity, followed by the interferometric reconstruction used to estimate signal direction.

\subsubsection{Signal Strength and Impulsivity}\label{sec:signalness}

Signal strength is quantified using the signal-to-noise ratio (SNR), defined as
\begin{align}
    \mathrm{SNR} = \frac{V_{\mathrm{pp}}/2}{{\mathrm{RMS}}}~,
    \label{eq}
\end{align}
where $V_{\mathrm{pp}}$ is the largest peak-to-peak voltage found over a rolling $20~\mathrm{ns}$ window over the waveform and ${\mathrm{RMS}}$ is the root-mean-square voltage measured in a noise-only portion of the waveform.

The SNR measures the amplitude of a signal relative to the thermal-noise baseline but does not describe its temporal structure. Impulsivity provides a complementary measure of how strongly the waveform power is concentrated in a short-duration pulse.
The impulsivity-related variables are inspired by ANITA~\cite{ANITA:2018impulsivity},
and are described in detail in Appendix~\ref{app:lda}.

Several signal-strength and impulsivity variables are also calculated using a coherently summed waveform (CSW).
The CSW is constructed separately for each polarization by time-aligning and summing the corresponding antenna waveforms using the relative time delays expected for the reconstructed direction.
Signals that arrive from the reconstructed direction add constructively, while uncorrelated thermal noise adds incoherently, improving the separation between signal-like and thermal-like events.

\begin{figure}
    \centering
    \includegraphics[width=\linewidth]{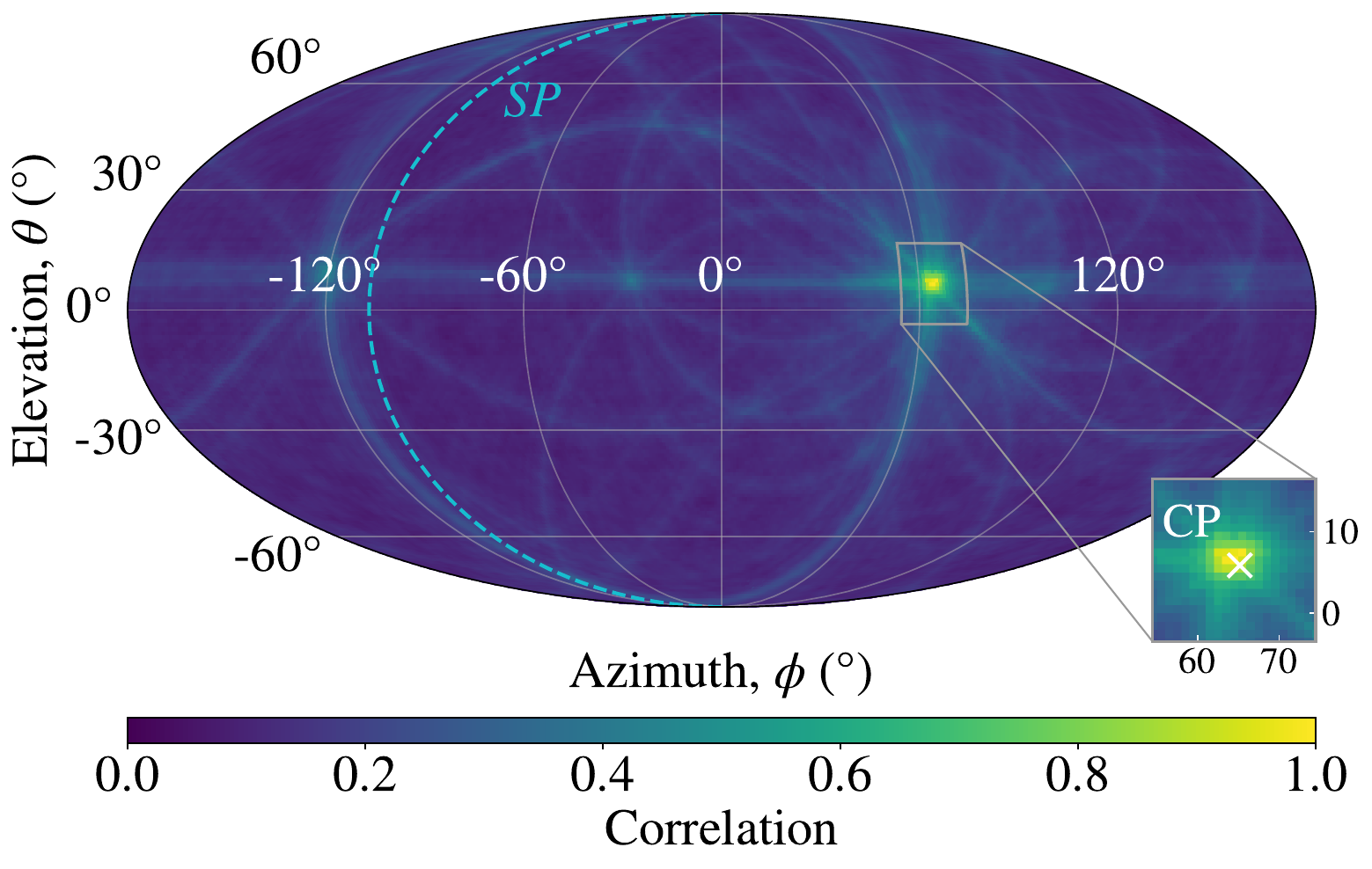}
    \caption{Interferometric skymap for the calibration pulser event shown in Fig.~\ref{fig:example_cp_event}. The map presents the average cross-correlation among VPol channels assuming an emission from the calibration pulser radius. The expected direction of the local calibration pulser (CP, white cross) is indicated, along with the South Pole direction (SP, cyan dashed line). \label{fig:skymap_example}
    }
\end{figure}

\subsubsection{Interferometric Reconstruction}\label{subsubsec:interferometric_reconstruction}

Interferometric reconstruction is used to estimate the direction to the signal emission point and is a key observable for separating in-ice neutrino candidates from backgrounds.
The interferometric reconstruction technique cross-correlates the voltage traces $V(t)$ recorded by pairs of channels, building on the methods used in previous ARA analyses~\cite{ARA:2014fyf,ARA:2019wcf,ARA:2022rwq}.
For each antenna pair $(i,j)$, the normalized cross-correlation is defined as

\begin{align}
    C_{ij}(\tau) = \frac{\int V_i(t)\,V_j(t+\tau)\,dt}{\mathrm{RMS}_i\,\mathrm{RMS}_j}~,
\end{align}
where $\tau$ is the relative time delay between the channels. For standard stations, the Hilbert envelope of $C_{ij}(\tau)$ is used when constructing the interferometric map, since this reduces sensitivity to the oscillations of the correlation function.

For a hypothesized emission point at a distance $R$ and direction $(\theta,\phi)$ relative to the station center (the average channel position), the expected propagation time delays between antennas are computed via ray tracing from the emission point to each antenna. 
We assume a depth-dependent model of the ice's index of refraction to perform ray tracing:

\begin{align}
    n(z) = n_\mathrm{ice} - (n_\mathrm{ice}-n_\mathrm{firn}) e^{-z/\Lambda}~,
    \label{eq:nofz}
\end{align}
where z is the depth below the surface, $n_\mathrm{ice}$ and $n_\mathrm{firn}$ are the ice's refractive indices in bulk and firn regions, respectively, and $\Lambda$ is the length scale of the transition between these two regions.
For this paper, we adopted the model used in~\cite{ARA:2022rwq} with $n_\mathrm{ice} = 1.780$, $n_\mathrm{firn} = 1.326$, and $\Lambda = 0.0202^{-1}$\,m~$\simeq 49.505$\,m.
This depth-dependent index means radio signals follow curved, rather than straight, trajectories as they propagate through the ice~\cite{ARA:2011testbed}.

The interferometric reconstruction of each event is performed on a grid of hypothesized emission points.
The grid uses a $1^\circ$ resolution in both elevation, $\theta\in[-90^\circ,90^\circ]$, and azimuth, $\phi\in[-180^\circ,180^\circ]$, and two radii: the distance to the local calibration pulser, typically $40-55$\,m, and a $300$\,m radius as a proxy for distant sources.\footnote{A $145$\,m radius is used for A1 due to its shallow depth.} Precomputed ray-tracing tables provide the expected time delay between antenna pairs for each grid point. The final reconstruction map is constructed as an SNR-weighted average over antenna pairs,
\begin{align}\label{eq:interferometric_map}
    C_{\mathrm{sky}}^{p}(\theta,\phi;R)
    = \sum_{\substack{i<j\\i,j\in p}}
    w_{ij}^{p}\,
    C_{ij}
    \!\left[\tau_{ij}(\theta,\phi;R)\right]~,
\end{align}
where $p$ is the antenna polarization, $\tau_{ij}(\theta,\phi;R)$ is the ray-traced relative time delay between antennas $i$ and $j$, and $w_{ij} \propto (\mathrm{SNR}_{i}-1)(\mathrm{SNR}_{j}-1)$ is the weight given to that antenna pair.
This weighting gives greater importance to antenna pairs containing stronger signals.
Reconstruction maps are calculated for each polarization and hypothesized emission radius for an event.
The reconstructed arrival direction is defined as the location of the global maximum of $C_{\mathrm{sky}}^{p}$.

Figure~\ref{fig:skymap_example} illustrates the directional reconstruction for a representative A1 calibration pulser event. The compact interferometric maximum is consistent with the expected direction to the calibration pulser, demonstrating the ability of the interferometric reconstruction to localize a near, in-ice source.
Tests with such local in-ice transmitters have demonstrated approximately degree-scale directional reconstruction~\cite{clark2019,lu2020,Hughes:2022thesis}, while measurements of deep, distant pulsers have validated the reconstruction across all five ARA stations~\cite{Allison:2017jpy,ARA:2023frx,Allison:2019rgg}, with comparable resolutions.

\begin{figure*}
    \centering
    \includegraphics[width=\linewidth]{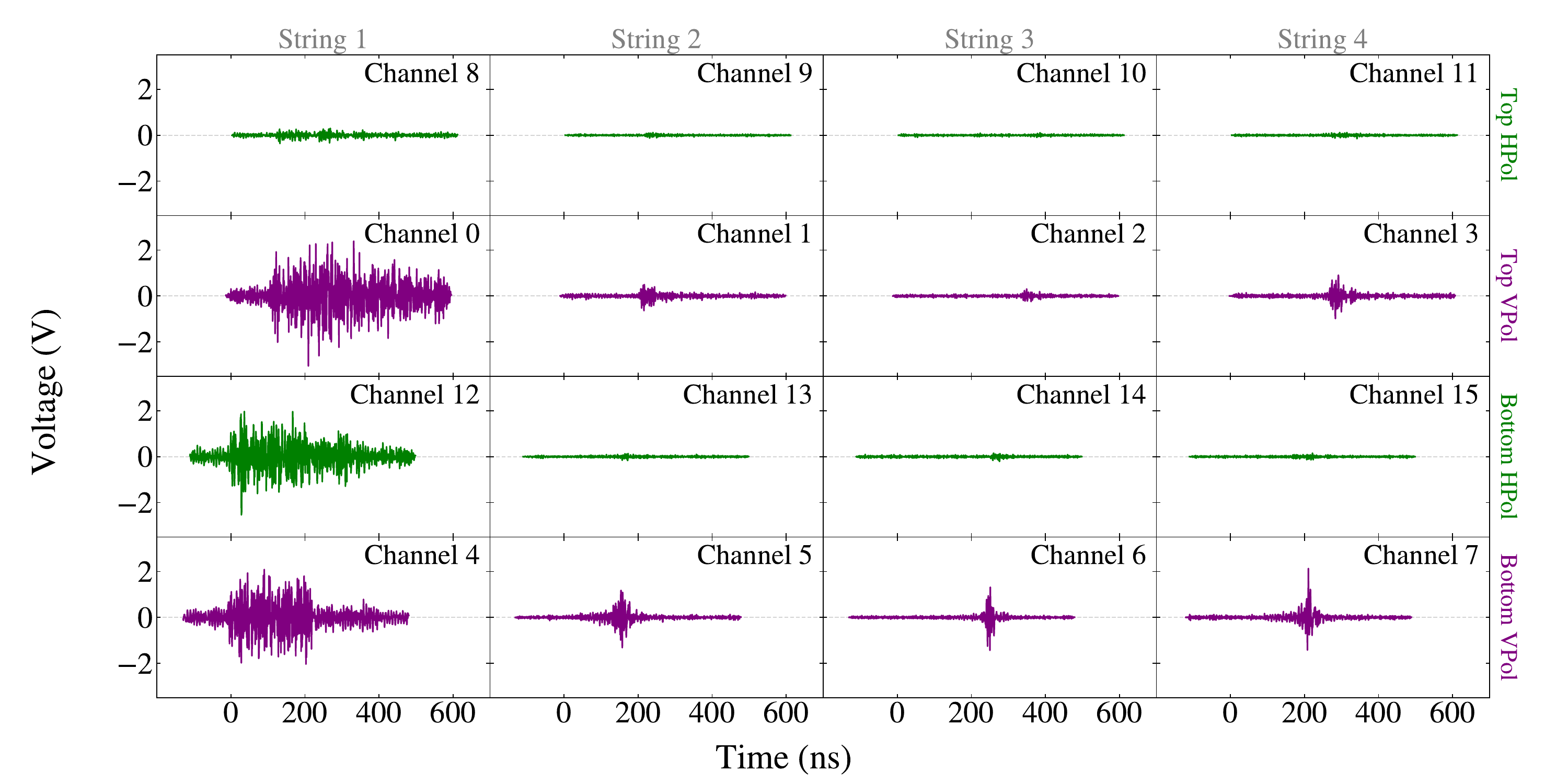} 
    \caption{Example voltage traces for an A5 event removed by the detector emission cuts. The four channels on the first string (channels 8, 0, 12, 4) exhibit sustained high RMS power across the readout window, while the remaining channels show no such anomalous power deposition.}
    \label{fig:detector_emission_event}
\end{figure*} 

\section{Event Selection}\label{sec:event_selection}

The event selection is applied in three stages, which are described in detail below. First, data quality cuts remove livetime and events that are not of analysis-quality. Second, station-level background cuts are applied to remove events associated with anthropogenic sources, cosmic rays, and thermal-like backgrounds. Finally, an array-wide time isolation cut is applied to remove any remaining anthropogenic backgrounds.

These selections were developed using the $10\%$~sample, leveraging the tail of the thermal-like population to set most thresholds. Background models for the surface background and thermal-like populations, as well as untagged calibration pulsers, were also developed from the $10\%$~sample. The thresholds for selections to remove the thermal-like and surface background populations were set by optimizing for $5\sigma$~discovery potential for a benchmark flux.
For this study, this flux was designed to saturate the existing experimental bounds.
The optimization of the cut thresholds and the expected background estimation are in Section~\ref{subsec:optimization}.

\subsection{Data Quality Cuts} \label{subsec:dataqualitycuts}

\subsubsection{Thermal-like Event Pre-Filter}\label{sec:fastfilter} 

Calculating the full set of analysis variables for every recorded event is computationally expensive. We therefore apply a pre-filter when analyzing the $100\%$~sample to remove events with timing patterns consistent with thermal noise before performing the full reconstruction.

The pre-filter uses two complementary measures of the arrival time differences between antennas. The first tests whether the measured delays are consistent with a common wavefront within each polarization~\cite{ARA:2019wcf}, while the second tests whether they are physically compatible with radio propagation through the ice. These variables are combined using a linear discriminant analysis trained on thermal-like data from the $10\%$~sample of events and simulated neutrino events.
The discriminant threshold is chosen such that it rejects no more than $10\%$ of the simulated neutrinos remaining after all other station-level event selections (see Section~\ref{sec:event_selection}). This is sufficient to reduce the data volume by approximately $60\%$ array-wide.

\subsubsection{Livetime Cuts} \label{subsubsec:livetimecuts}

Time periods in data taking which are contaminated by calibration campaigns, detector maintenance, or significant anthropogenic contamination are removed from the analysis.
Contaminated runs are identified both through experimental logbooks maintained by the collaboration and through an empirical metric.
The empirical measure is built by characterizing the distributions of maximum cross-correlations, as defined in Eq.~\eqref{eq:interferometric_map}.
Runs were removed if their distributions were significantly better described by a two-population (thermal and non-thermal) model, rather than a single population (thermal-only) model.
In addition to the removal of entire runs, the first minute of each data-taking run was excluded from the analysis.
This is done because the servo that controls the trigger threshold is not stable in the first minute after data taking has been initiated.

\subsubsection{Detector Emission Cuts}\label{subsubsec:detector_emission_cuts}

As discussed in a previous ARA search~\cite{ARA:2019wcf}, a deposition of power in the channels of a single string is sometimes observed which is inconsistent with our expectation of a neutrino event.
A prototypical event of this type is shown in Fig.~\ref{fig:detector_emission_event}.
This type of event is quite rare, as we only observed $7$~such events in a previous analysis of $\sim8$~station-years of data~\cite{ARA:2019wcf}.
The exact origin of this emission is actively being investigated, though it is currently hypothesized to be due to electrostatic discharge occurring near the antennas whose traces first exhibit a sudden increase in their RMS. 

Two complementary cuts were designed to remove this class of events. The first, originally developed in the previous search~\cite{ARA:2019wcf}, looks for excess power deposited in a single string relative to others and requires that the VPol and HPol channels agree in their reconstructed elevation to within $2.5^\circ$.
The second cut removes events that have a large RMS signal in a single channel that does not correlate well with the signal on other channels of the same polarization.

\begin{figure}
    \centering
    \includegraphics[width=\linewidth]{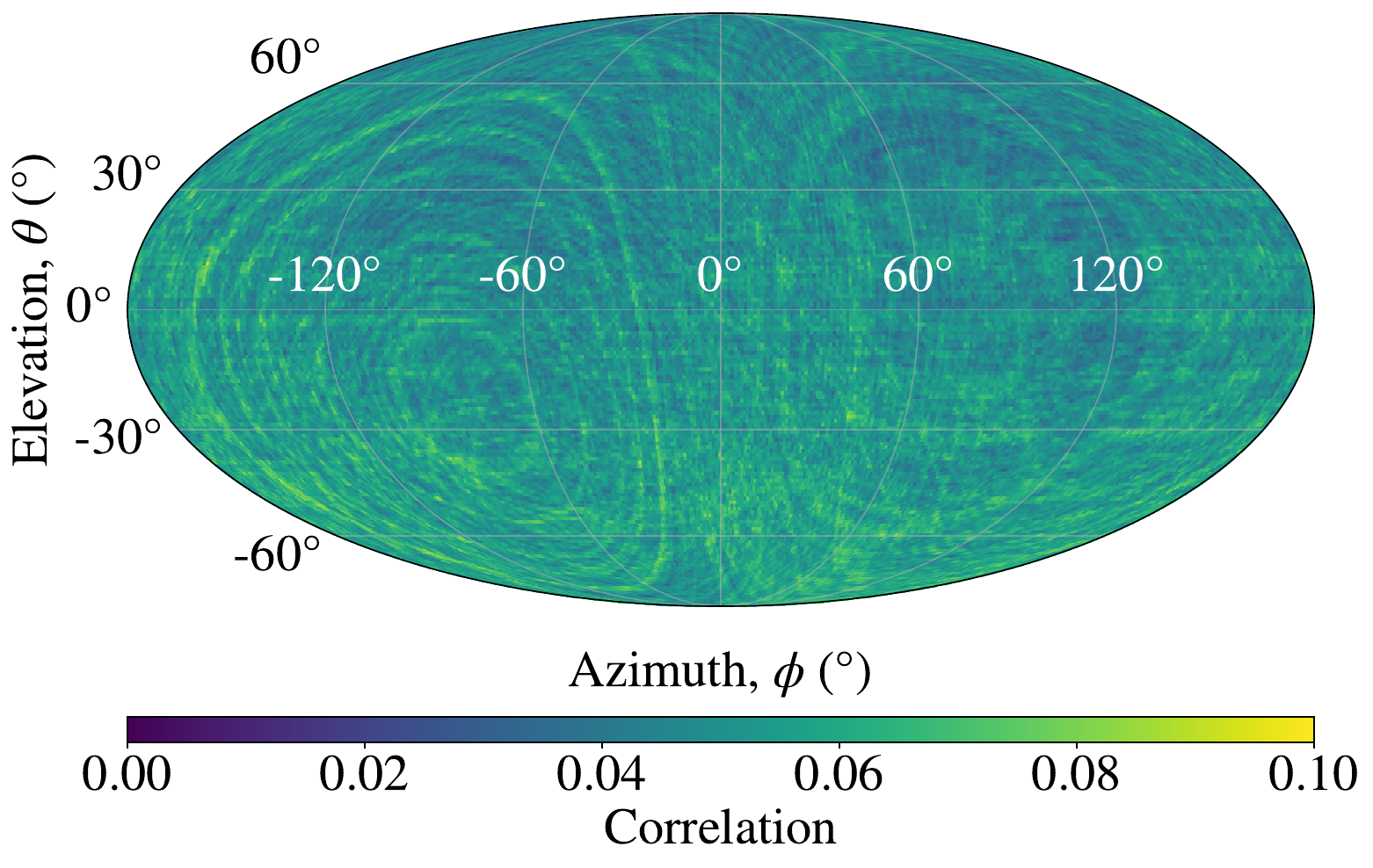} 
    \caption{Representative interferometric map for a low-contrast event. The correlation is distributed across broad and comparable structures rather than a single well-defined maximum. The reconstruction sphere is at the radius of the local calibration pulser.
    \label{fig:low_contrast_event}
    }
\end{figure}

\subsubsection{Reconstruction Quality Cut}

The final data quality cut we imposed was on the quality of an event's reconstruction.
We quantified the reconstruction quality by measuring the contrast of the point of maximum cross-correlation relative to the all-sky average in each event's interferometric reconstruction map. An example of a high-contrast event is shown in Fig.~\ref{fig:skymap_example}, while a low-contrast event is shown in Fig.~\ref{fig:low_contrast_event}.
Events that exhibited non-thermal features in their voltage traces (such as high SNR or high correlation) but had low-contrast interferometric reconstructions were removed from the analysis.

\subsection{Background Rejection Cuts}\label{subsec:background_cuts}

Once an analysis-quality dataset has been selected, we apply cuts to remove background events.
These cuts are applied at the station-level, except for an array-wide time isolation cut.
After these cuts, ideally the only remaining events should be neutrino candidates, along with a relatively small leakage of background events.

We apply cuts targeting different background populations.
First, geometric cuts are applied to remove untagged calibration pulsers and other known stationary sources.
Second, spatiotemporal cuts are applied to remove transient anthropogenic backgrounds, such as weather balloons, airplanes, and snowmobiles.
Third, surface cuts are applied to remove any remaining anthropogenic backgrounds, as well as cosmic ray backgrounds.
Finally, a thermal cut is applied to remove thermal-like backgrounds from the dataset.
These cuts are often not independent, instead complementing each other to suppress backgrounds.
After all station-level cuts, an array-wide time isolation cut is applied to remove any remnant anthropogenic backgrounds. Table~\ref{tab:cut_efficiencies} quantifies each cut's efficiency and its overlap with others.

\begin{table}
    \centering
    \begin{tabularx}{\columnwidth}{>{\centering\arraybackslash}X|S[table-format=3.1]|S[table-format=3.1]|S[table-format=3.1]|S[table-format=3.1]}
            \toprule
            \textbf{Cut Name} & \textbf{$b$ (\%)} & \textbf{$b_\mathrm{NT}$ (\%)} & \textbf{$\epsilon$ (\%)} & \textbf{$\epsilon_\mathrm{unique}$ (\%)} \\
            \midrule
            Event Pre-Filter & 62.2 & 1.2 & 99.0 & 100.0 \\
            Detector Emission & 0.0 & 2.7 & 99.5 & 99.5 \\
            Reco. Quality & 0.0 & 53.1 & 94.1 & 97.4 \\
            Geometry & 4.2 & 19.8 & 96.8 & 97.7 \\
            Spatiotemporal & 0.0 & 19.7 & 100.0 & 100.0 \\
            Surface & 15.0 & 100.0 & 56.1 & 69.7 \\
            Thermal Background & 100.0 & \multicolumn{1}{c|}{N/A} & 85.2 & 92.5 \\
            Time Isolation Cut & \multicolumn{1}{c|}{N/A} & \multicolumn{1}{c|}{N/A} & 99.9 & 99.9 \\
            \midrule
            Total & 100.0 & 100.0 & 42.8 & \multicolumn{1}{c}{N/A} \\
            \bottomrule
    \end{tabularx}
    \caption{
        \label{tab:cut_efficiencies}
        Summary of array-wide background rejection (based on $10\%$ sample data) and signal efficiencies for the event selection used in this analysis after livetime cuts. Simulation efficiencies are based on an $E^{-1}$ spectrum and are exposure-averaged over all array-wide configurations. $b$ is the fraction of data removed by each cut after all earlier cuts are applied. $b_\mathrm{NT}$ is the fraction of non-thermal-like events (selected via a threshold on the average VPol SNR) removed by each cut after all earlier cuts are applied. $\epsilon$ is the portion of simulated neutrinos that pass each cut array-wide after all earlier cuts have been applied. The pre-filter efficiency reported here is evaluated before applying the remaining selection cuts, whereas the rejection of no more than $10\%$ of simulated neutrinos described in Section~\ref{sec:fastfilter} is evaluated after applying all other event selections. $1-\epsilon_\mathrm{unique}$ is the portion of all simulated neutrinos that are uniquely removed by that cut array-wide.
    }
\end{table}

\subsubsection{Geometry Cut}\label{subsubsec:geometry}

\begin{figure}
    \centering
    \includegraphics[width=\linewidth]{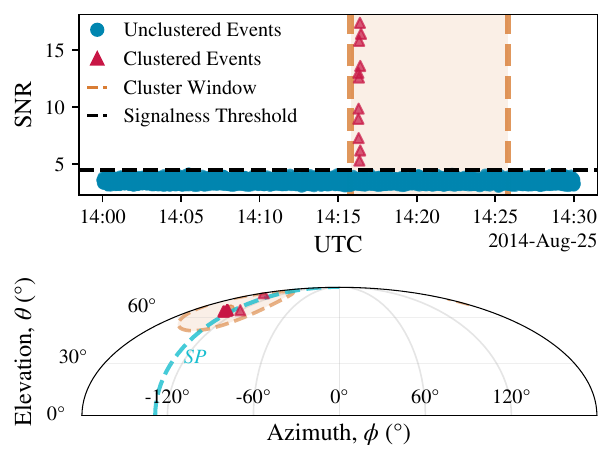} 
    \caption{
    A cluster in A3 identified by the spatiotemporal cut. Top: SNR as a function of time for triggered events, showing the time-clustered events. The signalness threshold and time window of the cut are indicated. Bottom: The spatial distribution of the above-threshold events in the selected time window, showing their directional clustering. The directional window and South Pole direction are indicated.} 
    \label{fig:stc_example}
\end{figure}

Known stationary sources of backgrounds are mitigated by removing all events that reconstruct toward the source direction.
The most prominent of these sources are each station's local calibration pulsers, since the signals they produce are, by design, Askaryan-like and reconstruct within the ice.
Most of these events are removed via an event-level tag assigned to calibration pulser events.
However, roughly $1$~in every $10^4$~calibration pulser events is not properly tagged and must be removed with a geometry cut.
We parametrized the angular distribution of tagged calibration pulsers with a Gaussian kernel density estimator (KDE), which is then used to set the cut's thresholds in order to control the leakage rate of untagged calibration pulsers into the signal region. 

An additional geometry cut was applied to A1 to mitigate the backgrounds from the direction of the South Pole. These backgrounds were more prevalent in this station due to its proximity to the South Pole and its shallow ${\sim}90$\,m depth. 

\subsubsection{Spatiotemporal Cut}\label{subsubsec:spatiotemporal}

Background sources that cannot be confined to one region of the sky or that are short-lived are removed via a spatiotemporal cut.
This cut selects non-thermal-like events that come from a common direction (solid angle window) within a short time window for removal.
The directional and temporal windows were chosen based on the typical scales of background events observed in the $10\%$~sample.
However, the threshold on the number of events required to form a cluster was chosen dynamically to control the rate of false clusters formed by upward fluctuations of the thermal-like backgrounds.
Only events identified as part of a spatiotemporal cluster were removed by this selection. An example of a spatiotemporal cluster removed by this cut is shown in Fig.~\ref{fig:stc_example}.
A detailed discussion of the construction of the spatiotemporal cut is provided in Appendix~\ref{app:spt}.

\subsubsection{Surface Cuts}\label{subsubsec:surface}

\begin{figure}[b]
    \centering
    \includegraphics[width=\linewidth]{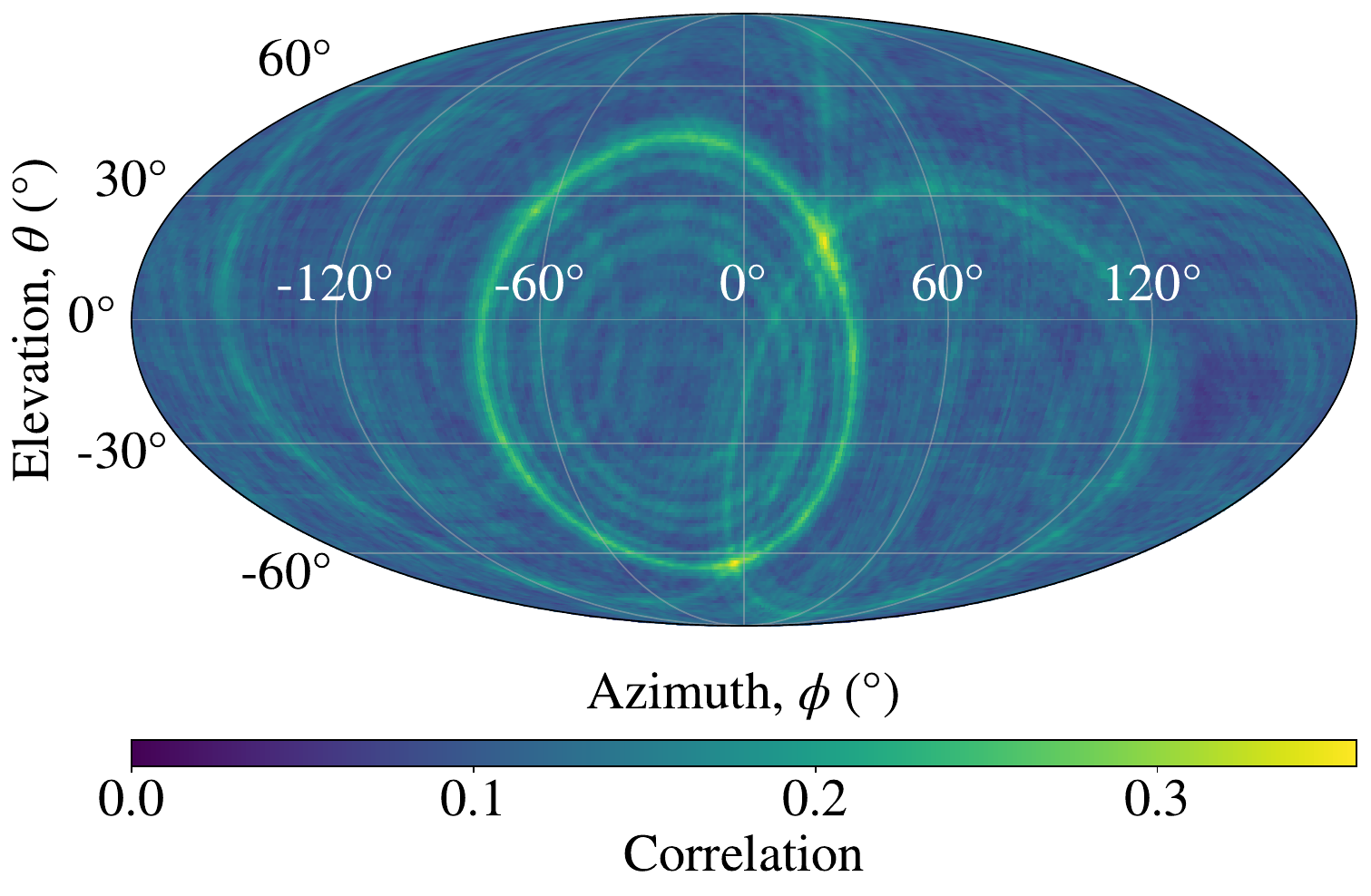} 
    \caption{
    Interferometric map of a representative surface-like event. The secondary peak and extended ring of correlated solutions illustrate that a surface-compatible interpretation can remain significant over a broad region of the map rather than only at the global maximum. The reconstruction sphere is at the radius of the local calibration pulser.
    \label{fig:surface_like_event}
    }
\end{figure}

The remaining non-thermal-like backgrounds in the dataset are dominated by anthropogenic backgrounds and cosmic ray events.
To mitigate these backgrounds, we apply two cuts to remove signals that reconstruct near or above the ice surface. 

\begin{figure}
    \centering
    \includegraphics[width=\linewidth]{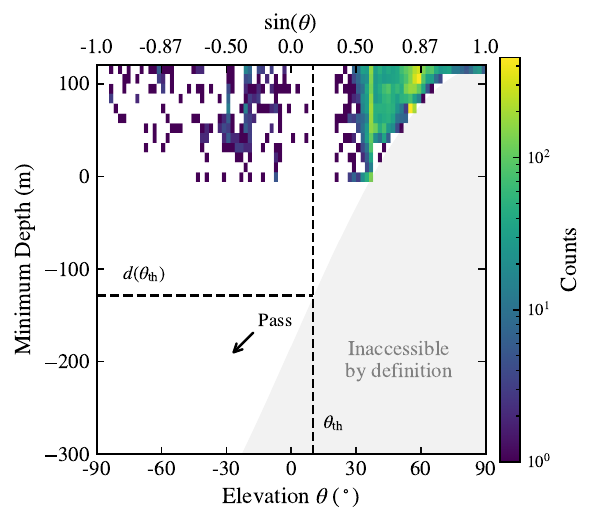}
    \caption{
    Distribution of A3 surface events in reconstructed elevation and minimum depth. Dashed lines mark the elevation cut threshold $\theta_{\rm th}$ and its corresponding depth $d(\theta_{\rm th})$. The grey region is inaccessible since the definition of minimum depth requires that $d \geq R \sin\theta +z_\mathrm{avg}$, where $R$ is the reconstruction radius and $z_\mathrm{avg}$ is the depth of the station-center. The passing signal region is indicated by an arrow.
    \label{fig:mindepth_vs_elevation}
    }
\end{figure}

The first of these cuts removes events that reconstruct towards the direction of the surface, above some threshold in elevation angle $\theta_\mathrm{th}$.
A second cut is applied to remove non-thermal-like events which have a secondary reconstruction solution in the direction of the surface, an example of which is shown in Fig.~\ref{fig:surface_like_event}. Such events often have secondary solutions, or interferometric rings of solutions, which are compatible with a surface origin.
These solutions are identified through a variable called ``minimum depth", defined as the shallowest depth at which the interferometric correlation reaches at least $60\%$ of its global maximum.
Events are removed if this depth is shallower than the depth corresponding to $\theta_\mathrm{th}$ at a fixed station-dependent radius. Figure~\ref{fig:mindepth_vs_elevation} shows an example distribution of minimum depth and reconstructed elevation for the $10\%$~sample dataset of A3, together with the thresholds for both cuts. Events in the top-left quadrant are determined to be ``surface-like'', in that their reported interferometric elevation is likely a misreconstruction. Further details are provided in Appendix~\ref{app:surface_background}.

A surface-dominated dataset was assembled, after removing mis-reconstructed events, to parametrize the surface background distribution in order to set the threshold for these cuts, $\theta_\mathrm{th}$.

\subsubsection{Thermal Background Cut}\label{subsubsec:thermal}

\begin{figure}[b]
    \centering
    \includegraphics[width=\linewidth]{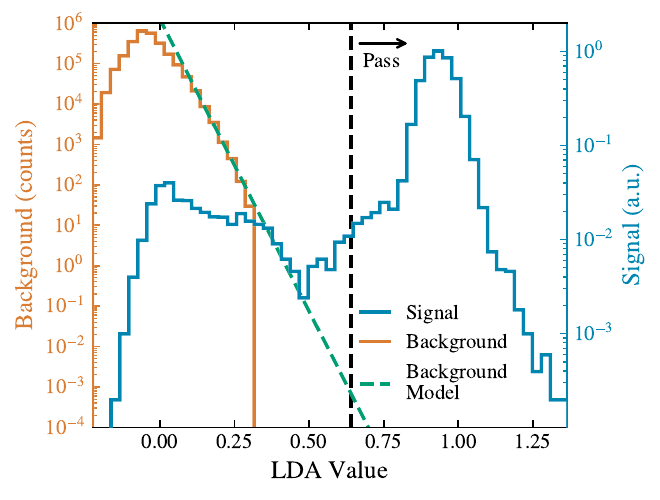}
    \caption{Thermal background and neutrino LDA score distributions for A3 during station-level configuration~2. The thermal background cut threshold for array-wide configuration~3 is shown (dashed black line). This array-wide configuration contains $80.51\%$~of the total A3 configuration~2 livetime. The simulated neutrino distribution (blue) is peak-normalized, while the $10\%$~thermal-dominated background distribution (orange) is scaled by the corresponding array-wide livetime fraction and by a factor of $10$~to represent the full dataset. The fit background model (dashed green line) is also shown.
    \label{fig:lda_performance}
    }
\end{figure}

After the data quality selections and previous background selections, the remaining data is dominated by thermal-like background events.
To remove this final population of events, we use a linear discriminant analysis (LDA) based cut.
The LDA is trained to maximize the separation of events from the $10\%$~sample and simulated neutrinos surviving all previous selections. An LDA is trained for each station-level configuration.  
Figure~\ref{fig:lda_performance} shows a representative example of the separation between signal and background populations achieved by the LDA.

The LDA score is formed via a weighted sum of analysis variables with weights chosen to maximize the separation of the signal and background data.
A subset of analysis variables was chosen for this purpose by removing strongly correlated variables to minimize redundant event characterizations.
The complete set of LDA input variables and their definitions is provided in Appendix~\ref{app:lda}. 

After training the LDA, the tail of the thermal-like background population is parametrized to set the threshold for the thermal background cut, $\sigma_\mathrm{th}$.
An example background fit is shown in Fig.~\ref{fig:lda_performance}.

\subsubsection{Time Isolation Cut}

A final array-wide event selection is applied after all station-level event selections to mitigate any remaining anthropogenic backgrounds. This selection comes in the form of a time isolation cut, motivated by the expectation that events from a small diffuse flux of neutrinos are unlikely to be observed close in time with one another.
Based on existing upper limits~\cite{ANITA:2019wyx,IceCubeCollaborationSS:2025jbi} (see Table~\ref{tab:eventcounts}), we assume, conservatively, that up to 15 neutrinos could be observed in our $10.6$~years of livetime.
Assuming these neutrinos are distributed randomly and uniformly throughout the livetime (which we take to be continuous for simplicity), we find $99.9\%$~of consecutive neutrinos have a temporal separation of at least $5.8$~hours.
We therefore require events to have this temporal separation, or greater, with one exception.
Any events recorded on separate stations that are within $5$~seconds of each other are selected for closer inspection before removal.\footnote{The $5$\,s window is orders of magnitude larger than both the light-travel time across the array and discrepancies between DAQ system clocks (which are known from external pulsing campaigns~\cite{Allison:2019rgg}).}
This is done in order to retain neutrinos which are observed by multiple stations~\cite{ARA:2026wjc}. Ultimately, this cut was not needed since only a single event survived to this stage.

\subsection{Optimization and Expected Background Estimate}\label{subsec:optimization}

The final event selection thresholds, those for the surface and thermal background cuts, are optimized globally over the full array livetime.
We optimize one thermal threshold $\sigma_\mathrm{th}$ per array-wide configuration per station (active in that configuration, see Fig.~\ref{fig:arraywide_configs}) and one surface threshold $\theta_\mathrm{th}$ per detector ($5$~standard detectors and the PA).
This results in $144$~thermal thresholds, collectively $\boldsymbol{\Sigma}$, and $6$~surface thresholds, collectively $\boldsymbol{\Theta}$, to be optimized.
This represents one of the largest suites of selection parameters simultaneously optimized in a radio-based diffuse neutrino search.

These thresholds are optimized for $5\sigma$~discovery potential of a flux $\phi(E) = \phi_0^{\mathrm{disc}}f(E)$, where $\phi_0^{\mathrm{disc}}$ is the flux normalization at an energy $E_0=10^{18.5}$~eV (the middle of the logarithmic energy range considered) and $f(E)$ parametrizes the spectral shape.
This shape function is defined such that $f(E_0)=1$.
For this analysis, we adopt the existing global upper limit on the UHE neutrino flux~\cite{ANITA:2019wyx,IceCubeCollaborationSS:2025jbi} as the spectral shape $f(E)$.
We then maximize the discovery potential by minimizing the flux normalization:
\begin{align}\label{eq:flux_minimization}
    \phi_0^{\mathrm{disc}}(\boldsymbol{\Sigma}, \boldsymbol{\Theta}) = \frac{\mathrm{FC}
    (n_\mathrm{disc}(B), B)
    }{\displaystyle \int f(E)\mathcal{E}(E)\,dE}~,
\end{align}
where $\mathrm{FC}(n, b)$ denotes the Feldman--Cousins $90\%$~confidence level (CL) upper limit~\cite{Feldman:1997qc} given $n$ observed events on an expected background $b$ including the correction from~\cite{Roe:1998zi}, $B=B(\boldsymbol{\Sigma}, \boldsymbol{\Theta})$ is the total background rate, $n_\mathrm{disc}$ is the $5\sigma$~discovery threshold, $\mathcal{E} = \mathcal{E}(E, \boldsymbol{\Sigma}, \boldsymbol{\Theta})$ is the analysis-level exposure, and the integral is taken over true neutrino energy from $10^{16}$~eV to $10^{21}$~eV. 

We estimate the total expected background in the signal region for our event selection by considering the leakage of three backgrounds: untagged calibration pulsers, surface events, and thermal-like background events. The untagged calibration pulser leakage is controlled by the geometry cut thresholds. These were set to achieve an expected leakage of $2\times 10^{-8}$~untagged calibration pulsers per pulser mode across the full analysis livetime, given the parametrized angular distribution of tagged calibration pulsers described in Section~\ref{subsubsec:geometry} and the untagged calibration pulser rate of $10^{-4}$. Given the PA detector's improved angular reconstruction, the geometry cut threshold was separately tuned to achieve an expected leakage of $10^{-8}$~untagged calibration pulsers per detector configuration. After accounting for the overlapping livetime of the A5 and PA detectors, this results in a total expected untagged calibration pulser leakage of $b_\mathrm{CP} = (4.2 \pm 0.3) \times 10^{-7}$~events.
This leakage was tuned to be several orders of magnitude smaller than those from the surface and thermal-like populations, discussed below. This was done to minimize the number of parameters in the optimization. 

The expected surface event leakage was estimated by integrating the parametrized surface background distributions, described in Section~\ref{subsubsec:surface}, beyond the cut $\theta_\mathrm{th}$. The total expected surface leakage is then given by
\begin{align}
    b_\mathrm{surf}(\boldsymbol{\Sigma}, \boldsymbol{\Theta}) = \sum_s b_s(\theta_{\mathrm{th},s}) f_s(\boldsymbol{\Sigma})~,
    \label{eq:surface_background}
\end{align}
where the sum is over stations $s$, $b_s$ is the surface leakage for station $s$ over the full analysis livetime, and $f_s(\boldsymbol{\Sigma})$ is a correction factor accounting for the fraction of surface events on station $s$ surviving the thermal cut. 

The expected thermal-like background leakage was estimated by integrating the parametrized thermal-like background distributions, described in Section~\ref{subsubsec:thermal}, beyond the cut $\sigma_\mathrm{th}$. The total expected thermal leakage is then given by
\begin{align}
    b_\mathrm{therm}(\boldsymbol{\Sigma}, \boldsymbol{\Theta}) = \sum_{a,s,c}  w_{s,c,a} b_{s,c}(\sigma_{\mathrm{th},s,a})f_{s}(\theta_{\mathrm{th},s})~,
    \label{eq:thermal_background}
\end{align}
where the sum is over array-wide configurations $a$, stations $s$, and station-level configurations $c$. The weight $w_{s,c,a}$ is the fraction of station-level configuration $c$'s livetime which occurs during array-wide configuration $a$. The term $b_{s,c}$ is the thermal leakage rate of the station in configuration $c$ for threshold $\sigma_{\mathrm{th},s,a}$. Finally, $f_{s}(\theta_{\mathrm{th},s})$ is a correction factor accounting for the fraction of the sky removed by the $s$th station's surface cut.

The total background leakage is then calculated as

\begin{align}\label{eq:total_background}
    B(\boldsymbol{\Sigma}, \boldsymbol{\Theta}) = b_\mathrm{therm}(\boldsymbol{\Sigma}, \boldsymbol{\Theta}) + b_\mathrm{surf}(\boldsymbol{\Sigma}, \boldsymbol{\Theta})+ b_\mathrm{CP}~.
\end{align}

We define the $5\sigma$~discovery threshold $n_\mathrm{disc}$ as the minimum number of observed events required to reject the background-only null hypothesis at $5\sigma$~significance.
Under the background-only hypothesis, the probability of observing at most $n$ events is given by the Poisson cumulative distribution function
\begin{align}\label{eq:poisson_cdf}
    C(n;B) =
    \sum_{k=0}^{n} \frac{B^{k}e^{-B}}{k!}~,
\end{align}
where $B$ is the expected background and $n_\mathrm{disc}$ is the smallest integer such that
\begin{align}
    C(n_\mathrm{disc};B) \geq 1-\beta~,
\end{align}
where $\beta = 2.87 \times 10^{-7}$ is the one-sided tail probability corresponding to $5\sigma$~significance.

After optimization, the final expected surface and thermal background leakages were $b_\mathrm{surf} = 4.50^{+1.01}_{-0.85}  \times 10^{-2}$~events and $b_\mathrm{therm} = 8.65^{+0.44}_{-0.28}\times 10^{-2}$~events. This resulted in a total expected background of $B=0.13 \pm 0.01$~events over the full analysis livetime and a $5\sigma$~discovery threshold of $n_\mathrm{disc} = 4$.

\begin{figure}    
        \centering\includegraphics[width=\linewidth]{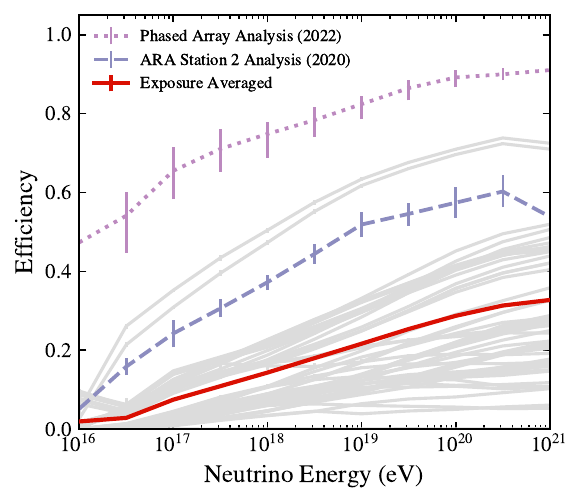}
        \caption{Exposure-averaged array-wide signal efficiency (red line) as a function of neutrino energy after the event selection. The signal efficiencies of individual array-wide configurations are also shown (gray lines); the two highest are array-wide configurations in which only the PA was active. For comparison, efficiencies from previous ARA analyses of the PA (purple dotted line) and A2/A3 (navy dashed line) are shown~\cite{ARA:2019wcf, ARA:2022rwq}.}
        \label{fig:signal_efficiency}
\end{figure}

The resulting array-wide energy-dependent signal efficiencies are shown in Fig.~\ref{fig:signal_efficiency},
with the efficiencies for each cut in Table~\ref{tab:cut_efficiencies}.
Studying the table, the least efficient stage of the event selection is the surface rejection cut.
This is partially because radio waves follow curved trajectories,
and can appear as downgoing signals when they in fact arise from within the ice.
The most efficient stage is the spatiotemporal cut, which, to our knowledge, has not previously been employed in analyses of in-ice Askaryan arrays. A detailed description of its estimated efficiency is provided in Appendix~\ref{app:signal_efficiency}.

Finally, our analysis-level exposure is calculated as
\begin{align}\label{eq:analysis_exposure}
    \mathcal{E}(E, \boldsymbol{\Sigma}, \boldsymbol{\Theta}) =
    \sum_a T_a \mathcal{A}_a(E) \epsilon_a(E,\boldsymbol{\Sigma}, \boldsymbol{\Theta})~,
\end{align}
where the sum is over array-wide configurations $a$, $T_a$ is the livetime of that array-wide configuration, $\mathcal{A}_a$ is its array-wide trigger-level acceptance, and $\epsilon_a(E)$ is the array-wide signal efficiency of the event selection. 

\section{Uncertainties}\label{sec:uncertainties}

\subsection{Statistical Uncertainties \label{sec:statistical}}

We estimated the statistical uncertainty on the expected background leakage by considering statistical uncertainties in the expected calibration pulser, thermal-like, and surface background leakages. These leakages and their statistical uncertainties are reported in Section~\ref{subsec:optimization}. In this Section, we detail how these uncertainties were calculated. 

For $b_\mathrm{therm}$, we perform pseudoexperiments to construct the background leakage distribution. For this, we sample the exponential tail's parameters from a bivariate Gaussian distribution whose covariance matrix is given by the covariance on the nominal fit parameters discussed in Sec.~\ref{subsubsec:thermal}. For each realization, we calculate the expected background leakage for the sampled parameters by integrating the corresponding exponential tail fit beyond the optimized LDA cut value. The error on $b_\mathrm{therm,s,c,a}$ was then calculated as the absolute difference between the nominal leakage and the 16th/84th percentiles of the resulting distribution. These uncertainties were added in quadrature across all active stations and array-wide configurations to obtain the total statistical uncertainty on $b_\mathrm{therm}$.

Bootstrapping was used to find the uncertainties on $b_\mathrm{surf}$ and $b_\mathrm{CP}$, since these backgrounds were modeled using Gaussian KDEs. Each station's surface and calibration pulser distributions were sampled with replacement to generate a bootstrap sample containing the same number of events as the original dataset. Each bootstrap sample was modeled with a Gaussian KDE in the same manner as the nominal distribution. The expected leakage from each sample was then computed by integrating the resulting KDE over the signal region. As with the thermal-like background, the statistical error was computed as the absolute difference between the nominal leakage and the 16th/84th percentiles of the bootstrapping leakage distribution. The uncertainties for each station were summed in quadrature to find the total statistical uncertainty on $b_\mathrm{surf}$ and $b_\mathrm{CP}$.

\begin{figure}
        \centering\includegraphics[width=\linewidth]{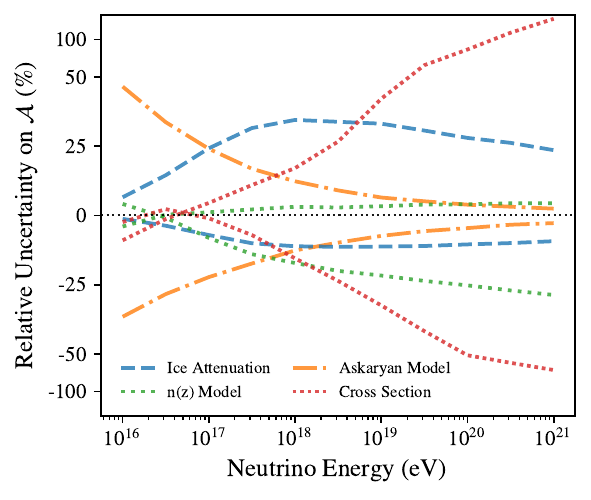}
        \caption{Relative uncertainty on the acceptance for array-wide configuration~40 for each source of systematic uncertainty. The vertical axis is linear within $\pm 50\%$ and logarithmic beyond this range.}
        \label{fig:systematics}
\end{figure}

\subsection{Systematic Uncertainties \label{sec:systematics}}

We evaluated the systematic uncertainties on the acceptance and on the background leakage established in Section~\ref{subsec:optimization}.
Four sources of systematic uncertainty were investigated for the calculation of ARA's acceptance: the refractive index model, the radio-wavelength attenuation model, the Askaryan emission model, and the neutrino cross section. The effects of each of these sources of uncertainty are shown in Fig.~\ref{fig:systematics} and are discussed in detail below.
Overall, these result in a systematic uncertainty of $+40\%/-29\%$ on the acceptance at $1$\,EeV.
Additionally, we considered two sources of systematic error for background leakage: the refractive index model and the phase response used during waveform pre-processing.
We calculate the systematic error on the expected background estimate to be $+0.05/-0.06$~events.

We follow largely the same systematics evaluation as the previous diffuse analyses~\cite{ARA:2019wcf,ARA:2022rwq} by calculating the change in the total acceptance under systematic variations relative to nominal.
The nominal index of refraction model, $n(z)$, is defined in Eq.~\eqref{eq:nofz}.
The parameters of this model have been estimated over the past three decades using many different datasets, including time-of-flight measurements and ice core samples. These estimates all generally agree at the tens-of-percent level~\cite{ARA:2019wcf, ARA:2022rwq, ARA:2026wjc}.
We vary the three parameters to bracket the range of models in the literature.
Namely, the upper envelope is defined as $n_{\trm{ice}} = 1.788$, $n_{\trm{firn}} = 1.410$, $\Lambda=48.5$\,m, while the lower envelope is $n_{\trm{ice}} = 1.774$, $n_{\trm{firn}} = 1.166$, $\Lambda=64.5$\,m.

The nominal radio wavelength attenuation model is a frequency- and temperature-dependent model based on fits to data from IceCube, AMANDA, and Bogorodsky~\cite{ARA:2026wjc}.
The uncertainty on the model is taken from attenuation length measurements made by the ARA prototype detector~\cite{ARA:2011testbed}.
At $300$\,MHz, averaged across depths,
this amounts to an uncertainty on the attenuation length of about $+15\%/-8\%$.
The asymmetry in this uncertainty propagates to the error on the effective area, and is visible in Fig.~\ref{fig:systematics}.

The nominal Askaryan emission model comes from fitting a parametrized model of Askaryan radiation to microscopic shower simulations~\cite{hong2014, ARA:2026wjc}.
These disagree with the amplitude of the microscopic simulations by at most $12\%$~\cite{hong2014}, which we adopt as the systematic uncertainty on the signal amplitude in simulation.
Varying the Askaryan model has the largest effect at low energies since the electric field amplitudes, which scale linearly with neutrino energy, are smallest and nearest to the trigger thresholds.

The nominal neutrino-nucleon interaction cross section is taken from Connolly et al.~\cite{Connolly:2011vc}, which also provides error bands that we adopt directly.
The uncertainties on the cross section in~\cite{Connolly:2011vc} increase with the primary energy, reaching over $100\%$~at $10^{21}$~eV, which propagate directly onto the acceptance.
We point out that~\cite{Connolly:2011vc} has the largest uncertainties, by a factor of several, of any cross section model available in the literature~\cite{Weigel:2024gzh, Cooper-Sarkar:2011jtt}, and so the range adopted here is conservative.

For each of the systematics considered for the background leakage, we re-analyzed all data for a representative station-level configuration of A2 under each of the systematic variations separately, and recalculated the expected background leakage. The errors we report come from the differences between the leakage found under each systematic variation and the nominal leakage, which are then added in quadrature to find the total systematic error.
The uncertainties for the refractive index model are the same as described above for the acceptance calculation.
We find that varying the refractive model had a $+3\%/-46\%$ impact on the expected background.
For the phase response, the nominal model is an $S$-parameter simulation of the ARA signal chain.
To assess the systematic error, we replace the simulated phase response with one measured in the laboratory.\footnote{The simulated phase response is used as nominal due to a lack of available measurements for antennas at each station and to maintain consistency with past analyses~\cite{ARA:2011testbed}.}
We find that this increases the expected background estimate by $+41\%$.
The total uncertainty on the expected background, arising from systematic uncertainties, is then $+0.05/-0.06$~events.

\section{Results}\label{sec:results}

After the event selection thresholds were established using the $10\%$~sample, the full dataset was unblinded. The unblinding process began with compatibility cross-checks to ensure that the thermal and surface background models developed with the $10\%$~sample were consistent with those background populations in the full dataset. After these compatibility cross-checks were completed, the station-level event selections were applied to the full dataset before the final time isolation cut was applied to the remaining events across all stations. However, as discussed below, this final cut did not remove any events from the analysis.

\begin{figure}        
        \centering
        \includegraphics[width=\linewidth]{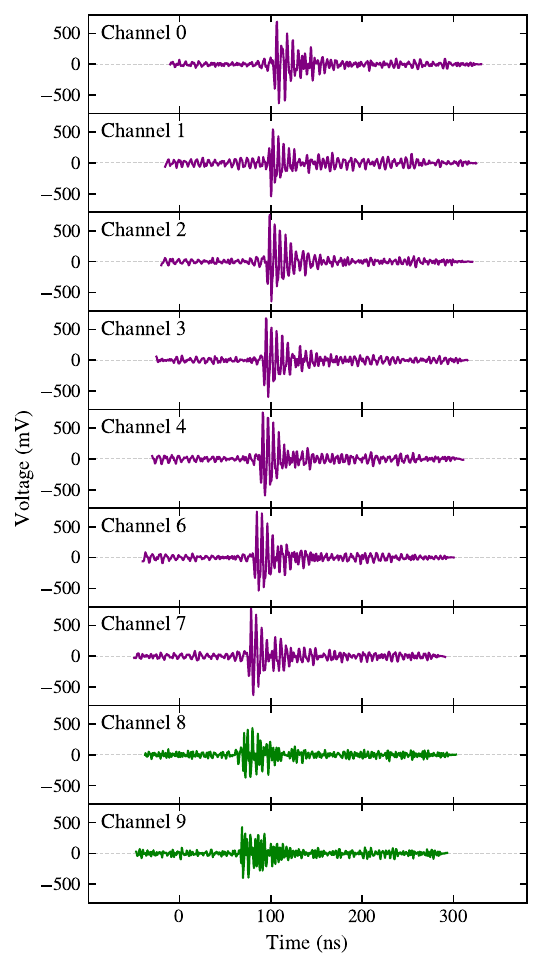}
        \caption{Waveform of the passing PA event in each VPol (purple) and HPol (green) channel. Channel number increases with depth. Channel 5 is omitted since it was not instrumented at the time the event was recorded. This event was observed simultaneously with the A5 event shown in Fig.~\ref{fig:detector_emission_event}.}
        \label{fig:passing_event_trace}
\end{figure}

\begin{figure}
        \centering\includegraphics[width=\linewidth]{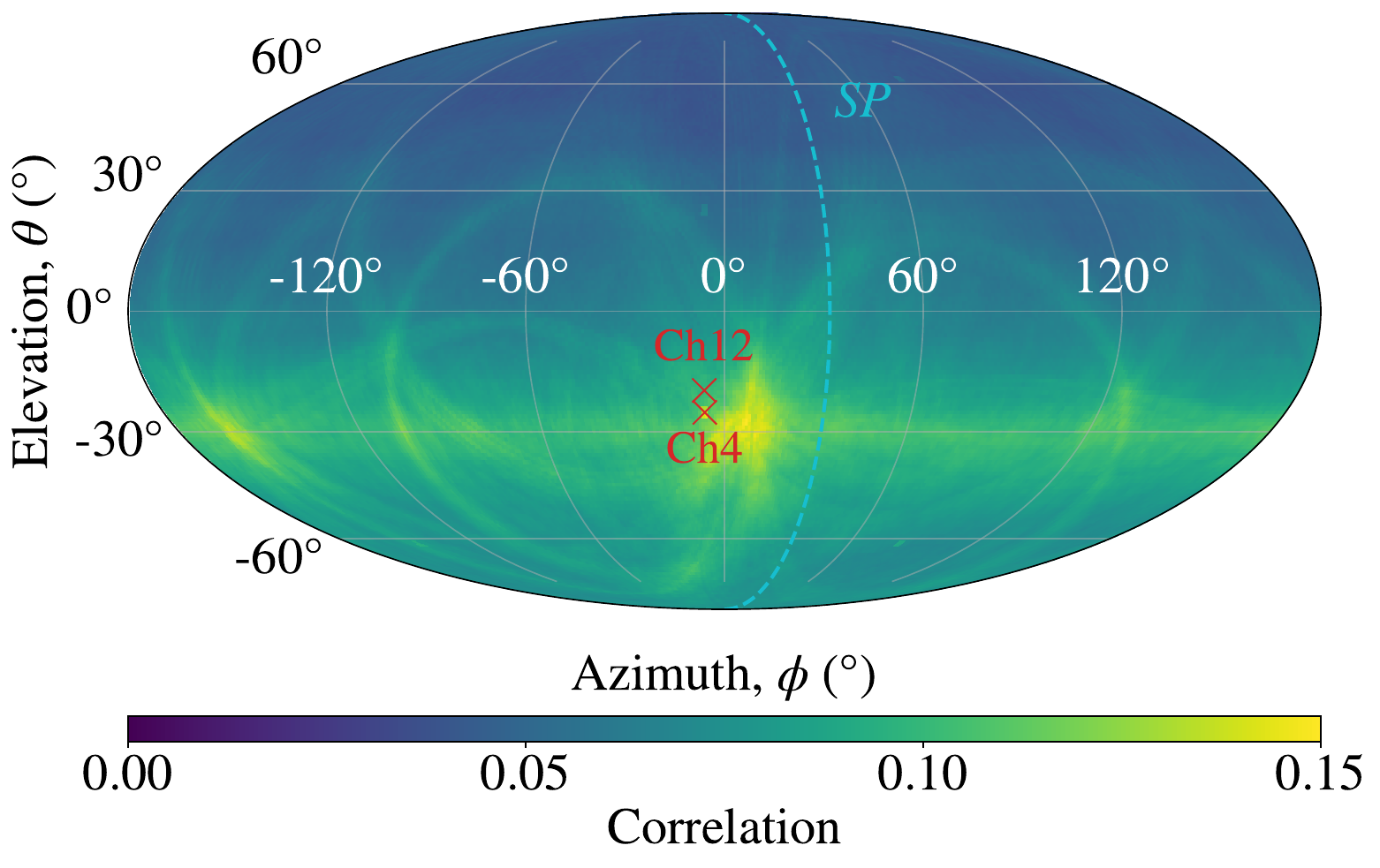}
        \caption{Joint A5-PA reconstruction map of the passing PA event assuming emission from a radius of $32.93$~m, corresponding to the distance to channels 4 and 12 on A5 (red crosses). The direction to South Pole (SP, dashed cyan line) is shown for reference. The reconstruction profiles over an unknown time offset between the A5 and PA channels, choosing the offset which maximizes the correlation in each direction.} 
        \label{fig:passing_event_joint_reco}
\end{figure}

One event survived the cuts as they were defined on the $10\%$~sample.
This event was recorded by the PA and was determined to be a background.
The event, shown in Fig.~\ref{fig:passing_event_trace}, is at first quite promising: it is impulsive, with an SNR of $\sim12$ in both polarizations, and a reconstructed elevation of $-35^\circ$.
However, further investigation revealed that a simultaneous event, shown in Fig.~\ref{fig:detector_emission_event}, was recorded by the co-located A5 detector.
This A5 event was removed from the analysis by the detector emission cuts (see Section~\ref{subsubsec:detector_emission_cuts}) due to anomalously large RMS traces on some channels.
The PA did not include detector emission cuts because such events were not observed in the $10\%$~sample.
For this reason, the possibility of the PA recording an anomalous emission from A5 was not considered.
A detailed reconstruction of the A5 event supported a local emission hypothesis, specifically near channels $4$ and $12$ (the channels showing the earliest large RMS signals in their traces). Both a PA-only and a joint A5-PA reconstruction, shown in Fig.~\ref{fig:passing_event_joint_reco}, of the passing event were consistent with this hypothesis.
Reconstruction maps for alternative source distance hypotheses can be found in Appendix~\ref{app:alt_recos}.
Therefore, the passing event on PA appears to have been caused by a detector emission event on A5, which generated an impulsive signal (presumably from channels $4$ and $12$) and triggered the PA detector.
An additional event selection was made on the PA dataset \textit{a~posteriori} to remove all events within $0.1$\,s of A5 events excluded by the detector emission cuts.
This cut resulted in an additional $21.9$\,s of livetime being removed from the PA dataset, having a negligible impact on the overall sensitivity.
After applying this cut, there are zero events remaining in the selection.

\begin{figure}
    \centering
    \includegraphics[width=\linewidth]{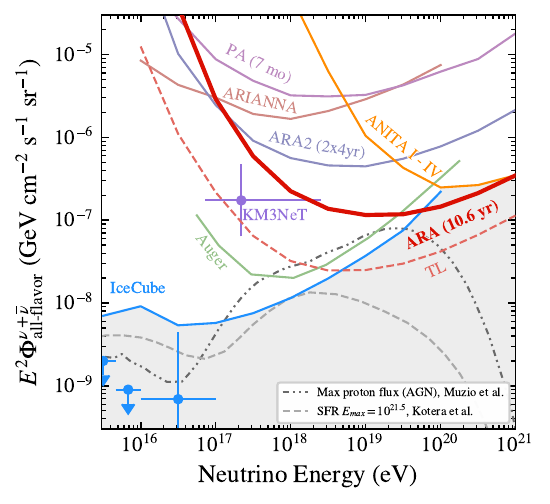}
    \caption{
        The $90\%$ CL upper limit on the all-flavor diffuse UHE neutrino flux set by this analysis (thick red line). The trigger-level sensitivity (TL, dashed red line) of this analysis is also shown. 
        Previous constraints and measurements from IceCube~\cite{IceCubeCollaborationSS:2025jbi, IceCube:2025ewu}, KM3NeT~\cite{KM3NeT:2025npi}, Auger~\cite{PierreAuger:2023pjg}, ANITA I--IV~\cite{ANITA:2019wyx}, ARIANNA~\cite{Anker:2019rzo}, and earlier ARA analyses~\cite{ARA:2019wcf,ARA:2022rwq} are also shown (colored lines). The boundary of the grey region represents the global upper limit flux used in optimization. All upper limits are converted to equivalent all-flavor limits using decade-wide bins. Adjustments are not made for differences in the assumed spectral index within each bin. 
        A selection of UHE neutrino flux models~\cite{Kotera:2010yn,Muzio:2023skc} are presented (grey dashed lines). 
        \label{fig:sensitivity}
    }
\end{figure}

Having observed $0$~passing events on an expected background of $0.13\pm0.01\text{ (stat.)}^{+0.05}_{-0.06}\text{ (syst.)}$~events, we set the upper limit shown in Fig.~\ref{fig:sensitivity} on the diffuse neutrino flux. 
Table~\ref{tab:eventcounts} presents the expected number of neutrino detections by ARA at the trigger- and analysis-level. This shows that the established upper limit probes both the neutrino flux established by the KM3NeT observation in~\cite{KM3NeT:2025npi} and the trans-GZK neutrino flux model of~\cite{Muzio:2023skc}. The latter model can be excluded at the $90\%$~confidence level.

\begin{table}
    \centering
    \begin{tabularx}{\columnwidth}{>{\centering\arraybackslash}X|S[table-format=2.2]|S[table-format=1.2]|S[table-format=1.2]}
    \toprule
    \textbf{Flux Model} & \textbf{Trigger} & \textbf{Analysis} & \textbf{$\mathbf{\epsilon}$ (\%)}  \\
    \midrule
    Global UL \cite{IceCubeCollaborationSS:2025jbi, ANITA:2019wyx} &  19.2 & 5.3 & 27.6 \\
    Muzio~\cite{Muzio:2023skc} & 11.0 & 2.4 & 21.8 \\
    Kotera~\cite{Kotera:2010yn} & 1.9 & 0.4 & 21.1 \\
    100\% Protons~\cite{Ahlers:2012rz} & 0.8 & 0.2 & 25.0 \\
    Fang Pulsars~\cite{Fang:2013vla} & 1.2 & 0.2 & 16.7 \\
    KM3NeT~\cite{KM3NeT:2025npi} & 11.6 & 1.6 & 13.8 \\
    \bottomrule
    \end{tabularx}
    \caption{
        \label{tab:eventcounts}
        Expected number of events for several UHE neutrino flux models, shown at both trigger- and analysis-level, as well as the signal efficiency under each model. 
        The expected number of events for the KM3NeT flux is calculated assuming an $E^{-2}$ spectrum over the energy range encompassed by its error bars.
    }
\end{table}

\section{Discussion}\label{sec:discussion}

The upper limit resulting from this analysis is the most stringent on the diffuse all-flavor UHE neutrino flux from $50$\,EeV to $1000$\,EeV, as shown in Fig.~\ref{fig:sensitivity}.  
This result represents the first array-wide UHE neutrino search with an array of in-ice radio detectors and is based on the largest cumulative dataset collected by an in-ice radio array to date.
More generally, the limit demonstrates that radio-based searches are beginning to probe predicted UHE neutrino fluxes and provide world-leading sensitivity at the highest energies.
By optimizing the station- and configuration-dependent event selections together according to their contributions to the sensitivity of the full array, this work provides a foundation for analyses with larger arrays of inhomogeneous, autonomous radio detectors.
Increased exposure, improved analysis efficiency, and larger detectors are required to extend the sensitivity of UHE neutrino observatories to probe lower flux models.

Future analyses can build on this search in several ways, in particular to improve their analysis efficiencies. 
Refinements of the surface cut will yield the largest improvements in signal efficiency.
This begins with station design: stations that are far (at least $5-10$\,km) from anthropogenic sources and which deploy antennas deep into the ice will inherently have smaller surface background populations. Further, densely instrumented strings with a high degree of symmetry, as exemplified by the PA, more robustly reconstruct the elevation of surface backgrounds. This translates into a more effective suppression of their tails (particularly from mis-reconstructions) into the signal region. This is because dense, symmetric instrumentation offers multiple samplings of the electric field and redundant interferometric baselines.
This can be seen by comparing the PA with A1. At $3$\,km from South Pole and a relatively shallow $90$\,m depth, A1 had an elevation threshold of about $-7^{\circ}$ and a surface cut efficiency of about $16\%$~(when applied in isolation for an $E^{-1}$ spectrum).
The PA, in contrast, has an elevation threshold of about $20^{\circ}$ and a surface cut efficiency of $86\%$.
Beyond station design, more sophisticated surface cuts in analysis using time-, direction-, and SNR-dependent thresholds can potentially yield dividends.
Finally, the development of cosmic ray vetoes will allow for significantly less conservative surface cuts to be applied.

Additional gains are possible by developing techniques to further separate signals from thermal-like backgrounds, which dominate the dataset statistically and set the floor for the analysis' sensitivity to low-SNR neutrinos.
The correlation score with a matched filter template could potentially provide such gains~\cite{Kim:2023svg}.

Finally, future large Askaryan detector arrays will need to focus on developing and maintaining stations which are as homogeneous and stable as possible.
Searches using inhomogeneous stations are possible for moderate-sized arrays, as this work demonstrates.
However, for larger arrays, inhomogeneities create an array that becomes factorially more complex to analyze and to characterize accurately in simulation, and also multiplies the dimensionality of the optimization problem.
Uniform station design and electronics, combined with robust pre-deployment testing, will ensure that even searches on large arrays remain tractable without sacrificing sensitivity. 

The systematics evaluations also yield insights.
We evaluated the impact of systematic uncertainties on the expected background estimate and find them to be at roughly the $\pm 40\%$~level, with the upward uncertainty dominated by the signal chain systematic.
This suggests that uncertainties in the signal chain are a leading systematic on the expected background estimate, which motivates careful measurements and cataloging for future detectors.
On the calculated exposure, the uncertainties are overwhelmingly dominated by the cross section at high energies. However, we note that those adopted here are several times larger than others in the literature (e.g.\ from~\cite{Cooper-Sarkar:2011jtt}).
Further theoretical work to reconcile these uncertainties will be key for interpreting future detections and inferring flux measurements.
Next-generation detectors would profit from renewed campaigns to measure the attenuation length and refraction index down to the few-percent level should the cross section uncertainties indeed be much smaller than the $20\%$~scale, as indicated by other calculations.

\section{Conclusion}\label{sec:conclusion}

We have conducted the first array-wide search for diffuse ultrahigh energy neutrinos with the Askaryan Radio Array (ARA) --- the first such search performed with any in-ice Askaryan detector array. This search combines data recorded between January 2013 and December 2023 by all five ARA stations, including the phased array (PA) detector at A5, corresponding to $25.9$\,station-years of analysis-quality livetime collected over $10.6$\,years. It is also the first ARA UHE neutrino search to optimize the event selection for $5\sigma$ discovery potential and account for signals from secondary particles and events detectable by multiple stations. 

No neutrino candidates are observed, consistent with the expected background of $0.13\pm0.01\text{ (stat.)}^{+0.05}_{-0.06}\text{ (syst.)}$ events.
We therefore place a $90\%$~CL upper limit on the diffuse UHE neutrino flux, which is the strongest limit from $50$\,EeV to $1000$\,EeV.
The resulting limit presented here provides a complementary probe of the flux proposed in~\cite{KM3NeT:2025npi} and excludes the trans-GZK flux model of~\cite{Muzio:2023skc} at the $90\%$~CL.

This search demonstrates that data from different stations and time periods can be combined within a single search. The array-wide and time-dependent optimization improved the sensitivity, but it also increased the complexity of data analysis and detector simulations. Future Askaryan arrays can reduce this complexity through extensive pre-deployment testing and maintaining stable operating configurations.

This search also provides guidance for background rejection in future experiments. The absence of surviving anthropogenic events demonstrates the effectiveness of radio-based experiments in suppressing such backgrounds, even for stations located near anthropogenic activity. Further sensitivity improvements are expected from dedicated vetoes for cosmic rays and more sophisticated removal of surface backgrounds.

As the first array-wide search for diffuse neutrinos with in-ice Askaryan detectors, this work demonstrates the feasibility of neutrino searches with larger next-generation arrays, including the Radio Neutrino Observatory at  Greenland (RNO-G)~\cite{RNO-G:2020rmc} and IceCube-Gen2 Radio~\cite{IceCube-Gen2:2020qha}. The roadmap provided by this analysis will ensure that future UHE neutrino searches with ARA and other Askaryan detectors deepen the global UHE neutrino sensitivity and provide a path to identify the origin of UHE cosmic rays.

\begin{acknowledgments}
This analysis was conducted by A.~Bishop, B.~Clark, P.~Dasgupta, P.~Giri, A.~Machtay, M.~Muzio, A.~Salcedo-Gomez, and M.~F.~H.~Seikh.
\noindent
The ARA Collaboration is grateful for support from the National Science Foundation through Award No.~2013134 and~2310095.
The ARA Collaboration designed, constructed, and now operates the ARA detectors. 
We would like to thank IceCube, and specifically the winterovers, for the support in operating the detector. 
Data processing and calibration, Monte Carlo simulations of the detector and of theoretical models, and data analyses were performed by a large number
of collaboration members, who also discussed and approved the scientific results presented here. 
We are thankful to Antarctic Support Contractor staff, a Leidos unit for field support and enabling our work on the harshest continent. 
We thank the National Science Foundation (NSF) Office of Polar Programs and Physics Division for funding support.
We further thank the Taiwan National Science Council's Vanguard Program NSC 92-2628-M-002-09 and the Belgian F.R.S.-FNRS and FWO.
K.~Hughes thanks the NSF for support through the Graduate Research Fellowship Program Award No.~1746045. 
A.~Connolly thanks the NSF for Awards No.~1806923 and No.~2209588 and also acknowledges the Ohio Supercomputer Center. A.~Connolly and K.~Hughes would like to also thank the Center for Cosmology and AstroParticle Physics at the Ohio State University for funding, travel, and other support.
S.~A.~Wissel thanks the NSF for support through CAREER Award No.~2033500.
A.~Vieregg, C.~Deaconu, N.~Alden, and P.~Windischhofer thank the NSF for Award No.~2411662 and the Research Computing Center at the University of Chicago
for computing resources.
R.~Nichol thanks the Leverhulme Trust for their support. 
K.D.~de~Vries is supported by European Research Council under the European Union's Horizon research and innovation program (Grant Agreement 763 No.~805486). 
D.~Besson, I.~Kravchenko, and D.~Seckel thank the NSF for support through the IceCube EPSCoR Initiative (Award ID No.~2019597). 
M.S.~Muzio thanks the NSF for support through the MPS-Ascend Postdoctoral Fellowship under Award No.~2138121. 
A.~Bishop thanks the Belgian American Education Foundation for their Graduate Fellowship support.
\\
\end{acknowledgments}

\appendix 

\section{Discussion of Spatiotemporal Cut}\label{app:spt}

The spatiotemporal cut described in Section~\ref{subsubsec:spatiotemporal} has $4$ parameters: the signalness threshold $\mathscr{s}_\mathrm{th}$, the maximum cluster time window $\Delta t$, the maximum cluster directional (solid angle) window $\Delta \Omega$, and a cluster size threshold $N_\mathrm{th}$. 

We characterize an event's signalness $\mathscr{s}$ in terms of variables like SNR and impulsivity. The signalness threshold (or set of thresholds) is used to select events as candidates to form a cluster, and is set based on the high-signalness tail of the thermal-like population. The maximum time and directional windows are determined based on the typical scales of spatiotemporal clusters found in data, and follow an inverse relationship with each other, i.e.\ larger time windows correspond to smaller directional windows, and vice versa. For this search, we enforced this behavior by adopting the following relation between $\Delta t$ and $\Delta \Omega$:

\begin{align} \label{eq:time_solidangle_window}
    \Delta \Omega =
    \begin{cases}
        4\pi\,\mathrm{sr}, & \Delta t\le 1\,\text{s},\\[5pt]\displaystyle
        4\pi\left(\frac{\Delta t}{1\,\mathrm{s}}\right)^{-\alpha}\,\mathrm{sr}, & 1\,\text{s}<\Delta t<1\,\text{yr},\\[7pt]\displaystyle
        1\,\text{deg}^2, & \Delta t\ge 1\,\text{yr},
    \end{cases}
\end{align}
where $\alpha=\ln\left(4\pi\,\text{sr}/1\,\text{deg}^2\right)/\ln\left(1\,\mathrm{yr}/1\,\mathrm{s}\right) \simeq 0.616$. The limiting cases used in this relation were based on the rough angular resolution and trigger rate of the detector. The power-law scaling used in Eq.~\eqref{eq:time_solidangle_window} was chosen as a linear function between these limiting cases in log-space. These choices of limiting cases and functional forms were not optimized, and could be refined by future studies.

Finally, the cluster size threshold determines the minimum number of events required to form a cluster. This threshold is determined dynamically to suppress the number of false clusters formed by thermal-like events and incorporates anisotropies of these events. To do this efficiently, the number of false clusters expected in each run is controlled at the order-of-magnitude level. Namely, we require that the expected number of false spatiotemporal clusters be $\mathcal{O}(1)$. Noting that the spatial and temporal clustering conditions are independent, the expected number of spatiotemporal clusters with $k$ events is approximately given by

\begin{align}\label{eq:spt_dynamic_size}
    E[k] \simeq \sum_{r\geq k}^n N_{\Delta t}(r) P_\mathrm{time}(r) \left(1- e^{-\mu_\mathrm{spatial}(k;r)}\right)~,
\end{align}
where $n = f_\mathrm{trig}\Delta t$ is the maximum cluster size, $f_\mathrm{trig}$ is the trigger rate within the run, $N_{\Delta t}(r) = N-n+1 + \max(n-r, 0)$ is the number of distinct time windows in the run, $N$ is the total number of events in the run, and $P_\mathrm{time}(r)$ is the probability to find a time cluster of size $r$, modeled as a hypergeometric distribution:

\begin{align}
    P_\mathrm{time}(r) = \frac{ \binom{b}{r} \binom{N-b}{n-r}}{ \binom{N}{n} }~,
\end{align}
where $b$ is the total number of above-threshold events in the run. Finally, $\mu_\mathrm{spatial}(k;r)$ is the expected number of spatial clusters of size $k$ created from $r$ above-threshold events, so that the final term in Eq.~\eqref{eq:spt_dynamic_size} corresponds to the Poisson probability to find at least $1$~spatial cluster of size $k$. The expected number of spatial clusters of size $k$ is calculated by modeling spatial clustering as a binomial process:

\begin{align}
    \mu_\mathrm{spatial}(k;r) \simeq N_\mathrm{caps} {r \choose k} p_\mathrm{in}^k (1-p_\mathrm{in})^{r-k}~,
\end{align}
where $N_\mathrm{caps} = 4\pi/\Delta\Omega$ is the approximate number of independent spherical caps on the sky and $p_\mathrm{in}$ is the probability for an event to reconstruct within that cap. This direction-dependent probability is determined by assessing the distribution of thermal-like events in data. The spatiotemporal cluster size threshold $N_\mathrm{th}$ is then determined by finding the value of $k\geq 2$ such that $E[k] \sim 1$.

To make identification of spatiotemporal clusters computationally tractable, we only consider clusters within a single data-taking run (typically $3-6$~hours). Longer timescale clusters, which are inherently required to be directionally compact, are addressed by the geometry cut described in Section~\ref{subsubsec:geometry}. Clusters are formed algorithmically by identifying all events with $\mathscr{s}\geq \mathscr{s}_\mathrm{th}$, identifying the above-threshold events that are within a time $\Delta t$ of each other, and then identifying the subset of those events which are also contained within a spherical cap of size $\Delta \Omega$. Finally, this candidate cluster's size is compared to the dynamic threshold $N_\mathrm{th}$. If the candidate cluster is large enough, all of its constituent events are marked for removal. However, these events are still allowed to participate in the formation of subsequent clusters. 

\section{Spatiotemporal Cut Signal Efficiency}\label{app:signal_efficiency}

Whether a neutrino observed by a station would be cut by the spatiotemporal cut (see Appendix~\ref{app:spt} for details) depends on the time it arrives and its reconstructed direction $(\theta, \phi)$. If a spatiotemporal cluster of size $N\geq N_\mathrm{th}-1$ is formed by non-neutrino events overlapping the time and direction of an above-threshold neutrino, then the neutrino would be removed by the event selection since it would provide the critical event to surpass the threshold $N_\mathrm{th}$. 

Since the number of triggered neutrinos is expected to be small, we can then compute a data-driven estimate of the expected signal efficiency to above-threshold neutrinos reconstructing to a direction $(\theta,\phi)$ as  

\begin{align}
    \epsilon(\theta, \phi) = 1 - \frac{T_\mathrm{cut}(\theta,\phi)}{T_\mathrm{livetime}}~,
\end{align}
where $T_\mathrm{cut}(\theta,\phi)$ is the cumulative time that spatiotemporal clusters of size $N\geq N_\mathrm{th}-1$ cover the direction $(\theta, \phi)$ and $T_\mathrm{livetime}$ is the total livetime. In this analysis, we used a HEALPix pixelization~\cite{Gorski:2004by} of the sky to calculate the $T_\mathrm{cut}(\theta,\phi)$  from the data for each spatiotemporal cut on a station.

The total spatiotemporal cut efficiency is then calculated as

\begin{align}
    \epsilon_\mathrm{spatiotemporal} = \frac{\sum_i \eta_i w_i}{\sum_i w_i}~,
\end{align}
where the sum is over triggered simulated neutrino events, $w_i$ is the Monte Carlo weight of the $i$th event, and $\eta_i$ is a weight function encoding the expected outcome of the cut on event $i$:

\begin{align}
    \eta_i = 
    \begin{cases}
    \epsilon(\theta_i,\phi_i), & \mathscr{s}_i \geq \mathscr{s}_\mathrm{th} \\
    1, & \mathscr{s}_i < \mathscr{s}_\mathrm{th}
\end{cases}~.
\end{align}

\section{Surface Background Characterization}\label{app:surface_background}

As discussed in Section~\ref{subsubsec:surface}, a data-driven approach was used to remove backgrounds originating from the surface. These events are characterized using two complementary observables derived from the interferometric reconstruction described in Section~\ref{subsubsec:interferometric_reconstruction}: the reconstructed elevation angle of the correlation maximum and the ``minimum depth" variable. The minimum depth variable is used to identify events whose maximum correlation is below the surface region but whose interferometric map contains secondary solutions near the surface, as shown in Fig.~\ref{fig:surface_like_event} and Fig.~\ref{fig:mindepth_vs_elevation}.

The minimum depth of an event corresponds to the shallowest reconstructed elevation angle with a correlation value comparable to the maximum correlation value. For each VPol interferometric map, the correlation is projected onto the elevation axis by taking the maximum over all azimuths,

\begin{align}
    C_{\max}(\theta) = \max_{\phi} C(\theta,\phi)~.
\end{align}
The highest elevation angle satisfying

\begin{align}
    C_{\max}(\theta_{\min}) \geq 0.6~C_{\mathrm{max}}~,
\end{align}
where $C_\mathrm{max}$ is the global maximum correlation, is then converted to an equivalent depth via,

\begin{align}
    d(\theta) = R\sin\theta + z_{\mathrm{avg}}~,
\end{align}
where $R$ is the interferometric reconstruction radius and $z_{\mathrm{avg}}$ is the average depth of the station antennas, which is the center of the reconstruction sphere. Events with $d(\theta_\mathrm{min}) \geq d(\theta_\mathrm{th})$ are surface-like, where $d(\theta_\mathrm{th})$ is calculated for a sphere with radius of the calibration pulser for A1 and $300$\,m for A2-5 and the PA.

\begin{figure}
    \centering
    \includegraphics[width=\linewidth]{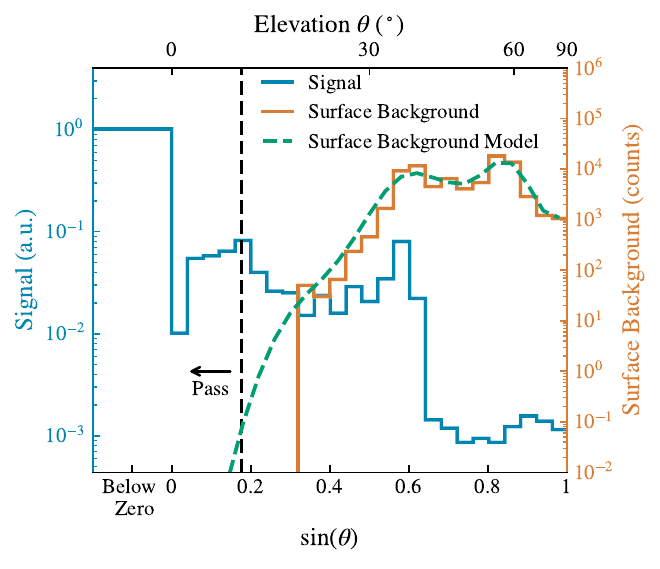}
    \caption{Reconstructed elevation distributions for the A3 surface-origin background sample and simulated neutrino signal for station-level configuration~2, along with the A3 surface cut threshold (dashed black line). The simulated neutrino signal (blue) is peak-normalized, while the $10\%$ surface-background distribution (orange) is scaled up by a factor of 10 to represent the full dataset. The KDE used to model the surface-background elevation distribution (green dashed curve) is also shown.}
    \label{fig:zenith_cut}
\end{figure}

To characterize the surface background population, a surface-dominated dataset was constructed for each station from triggered events in the $10\%$~sample that survive the data quality cuts described in Section~\ref{subsec:dataqualitycuts}. To have a sufficiently large dataset representative of these backgrounds, these datasets were constructed using runs that were otherwise removed in this analysis by the livetime cut described in Section~\ref{subsubsec:livetimecuts}. Runs that are known to include deep pulsing campaigns were excluded from this dataset. Further, to avoid including runs containing deep pulsing campaigns that were not tagged as such, runs with up-going events that were not ``surface-like'' (i.e.\ likely mis-reconstructions) were removed. After conducting the above cleaning of the surface-like dataset, a station-dependent SNR threshold was applied to suppress thermal-like events. 

\
\newcolumntype{Y}{>{\raggedright\arraybackslash}X}

\begin{table*}[htbp]
\centering
\label{tab:lda_variables}
\begin{tabularx}{0.8\textwidth}{@{}cY@{}}
\toprule
\# & Description \\
\midrule
1  & Average VPol channel SNR \\
2  & Average HPol channel SNR \\
3  & SNR of the coherently-summed VPol waveform (CSW) \\
4  & Root-power-ratio of the VPol CSW \\
5 & Ratio of peak-to-average VPol waveform power \\
6  & Maximum VPol interferometric correlation \\
7  & Maximum HPol interferometric correlation \\
8  & Contrast between maximum and mean VPol correlation, $C_{\mathrm{contrast}}$ \\
9  & Minimum depth: shallowest depth at which the VPol correlation reaches $60\%$ of its maximum \\
10  & Ratio of shallow-ice to global maximum VPol correlation, $R_{\mathrm{surf}}$ \\
11 & VPol impulsivity, $\mathcal{I}$ \\
12 & Fitted width parameter $B$ of $H_{\mathrm{tot}}$ \\
13 & $\chi^2$ of the impulse-plus-linear fit to the VPol impulsivity CDF, $\chi^2_{\trm{impulse}}$ \\
14 & Significance of the impulse-plus-linear fit over the linear-only fit, $\xi_{\trm{tot}}$ \\
\bottomrule
\end{tabularx}
\caption{LDA input variables used in the thermal background discriminator. \label{tab:lda_vars}}
\end{table*}

The surface background distributions for each station were used by the optimizer to set each elevation cut.
To compute the expected surface background leakage $b_s(\theta_{\mathrm{th},s})$ for station $s$, a Gaussian KDE constructed from the surface-dominated dataset (and scaled up by a factor of $10$ to represent the $100\%$ sample) is integrated below the cut threshold $\theta_{\mathrm{th},s}$. This KDE uses a bandwidth of $3^\circ$, chosen to conservatively account for uncertainties in the reconstructed signal arrival direction. The threshold $\theta_{\mathrm{th},s}$ was then optimized for each station to minimize the total expected surface background leakage in Eq.~\eqref{eq:surface_background}.

\section{Thermal Background Discriminator}
\label{app:lda}

The final cut of the event selection uses a Linear Discriminant Analysis (LDA)
to separate signals from thermal-like backgrounds.
We chose fourteen discriminating variables that were not strongly correlated and utilized them in the LDA.
In this Appendix, we describe each of these variables; they are summarized in Table~\ref{tab:lda_vars}.

\textit{Signal Strength Variables:} The first suite of variables we developed
focuses on quantifying signal strength.
For this, we calculate the average VPol channel SNR,
the average Hpol channel SNR,
the SNR of the coherently-summed waveform (CSW) in VPol channels,
the root-power-ratio (RPR) of the VPol CSW,
and the ratio of the peak-to-average VPol waveform power.

The RPR is defined as $\mathrm{RPR}=E_{\trm{max}}/\sigma_{E_{\trm{noise}}}$, where $E_{\trm{max}} = \max_j E_j$ is the maximum of the square root of a rolling $25$~ns integrated power average of the waveform, specifically:

\begin{equation}\label{eq:RPR}
    E_j = \sqrt{\frac{1}{n} \sum_{i=j}^{j+n} V_i^2}~,
\end{equation}
where $n$ is the number of samples in the $25$~ns window and $\sigma_{E_{\trm{noise}}}$ is the RMS value of $E_{j}$ in the noise-only portion of the waveform. This RPR variable has been used in previous ARA analyses~\cite{Allison:2015eky, ARA:2019wcf}, and emulates the power-integrated envelope used in the ARA trigger.

\textit{Reconstruction Variables:} The second suite of variables we developed
focus on reconstruction quality.
For this, we provide to the LDA the maximum VPol interferometric correlation and the maximum HPol interferometric correlation from the maps described in Section~\ref{subsubsec:interferometric_reconstruction}.
We additionally calculate the correlation contrast
$C_{\mathrm{contrast}}$.
We define this as the contrast between the maximum 
cross-correlation and the all-sky averaged 
cross-correlation in the VPol interferometric map:

\begin{align}
    C_{\mathrm{contrast}} = \frac{C_{\max}-\langle C\rangle}{\langle C\rangle}~,
\end{align}
where $\langle C\rangle$ is the mean correlation over the interferometric map.
$C_{\mathrm{contrast}}$ is designed to quantify the prominence of the peak.
We also calculate two variables sensitive to the surface origin of events.
The first is the ratio of the maximum VPol correlation in or above the top $10$~m of ice to the global maximum VPol correlation, defined as
\begin{align}
    R_{\mathrm{surf}} = \frac{C_{\mathrm{surf,max}}}{C_{\mathrm{global,max}}}~.
\end{align}
The second is the ``minimum depth'', the shallowest reconstructed depth at which the VPol correlation reaches at least $60\%$ of the maximum correlation, which is also described in Appendix~\ref{app:surface_background}.
Both attempt to quantify whether there is a surface- or near-surface-origin solution comparable to the global maximum.

\textit{Impulsivity Variables:} The third and final suite of variables
focuses on measuring the impulsiveness of an event, as this is an expected signature of Askaryan emission but not thermal-like noise.
We take inspiration from variables developed by the ANITA Collaboration~\cite{ANITA:2018impulsivity} and the prior ARA PA analysis~\cite{ARA:2022rwq}.
We start by taking the Hilbert envelope of the VPol CSW: $h(t)$.
We integrate outwards from the peak in the envelope, and normalize by the integral over the full envelope, to form a CDF-like variable of the Hilbert envelope, which we call $H(t)$
\begin{align}\label{eq:impulsivity_cdf}
    H(t) = \frac{1}{N} \int_{t_\mathrm{peak}-t/2}^{t_\mathrm{peak}+t/2} h(t') dt'~,
\end{align}
where $t_\mathrm{peak}$ is the time of the peak in $h(t)$, $N=\int_0^T h(t) dt$, and $T$ is the total waveform length.
Since traces may be of different lengths, we re-parametrize
the $H$ in terms of $t_{\trm{frac}}\equiv t/T$, the fractional time in the trace relative
to the peak in the envelope.
Meaning $t_{\trm{frac}}=0$ is the peak in the envelope,
while $t_{\trm{frac}}=1$ is when the end of the trace is reached.
This means that $H(t_{\trm{frac}}=0)=0$ and $H(t_{\trm{frac}}=1)=1$.
We provide to the LDA the impulsivity $\mathcal{I}$, defined as
\begin{equation}
    \mathcal{I} = 2 \langle H \rangle - 1~,
\end{equation}
where $\langle H \rangle$ is the time-average of $H$.
Since pure thermal noise has no impulsive time structure, $h_{\trm{noise}}(t_{\trm{frac}}) \propto \trm{constant}$ and so $H_{\trm{noise}}(t_{\mathrm{frac}}) \propto t_{\mathrm{frac}}$. Therefore, pure thermal noise has $\langle H \rangle \simeq 0.5$ and $\mathcal{I} \simeq 0$.
Meanwhile, an impulsive signal would generate a sharp bump at $t_{\trm{frac}}=0$,
which we generically parametrize as a Gaussian, so that $h_{\trm{impulse}}(t_{\trm{frac}}) \propto A e^{-t_{\trm{frac}}^2/B^2}$, where 
the parameters $A$ and $B$ control the amplitude and width of the impulse.
A perfect impulse would represent
a step function in $H$, and so 
$\langle H \rangle \simeq 1$ and $\mathcal{I} \simeq 1$.

\begin{figure}
        \centering\includegraphics[width=\linewidth]{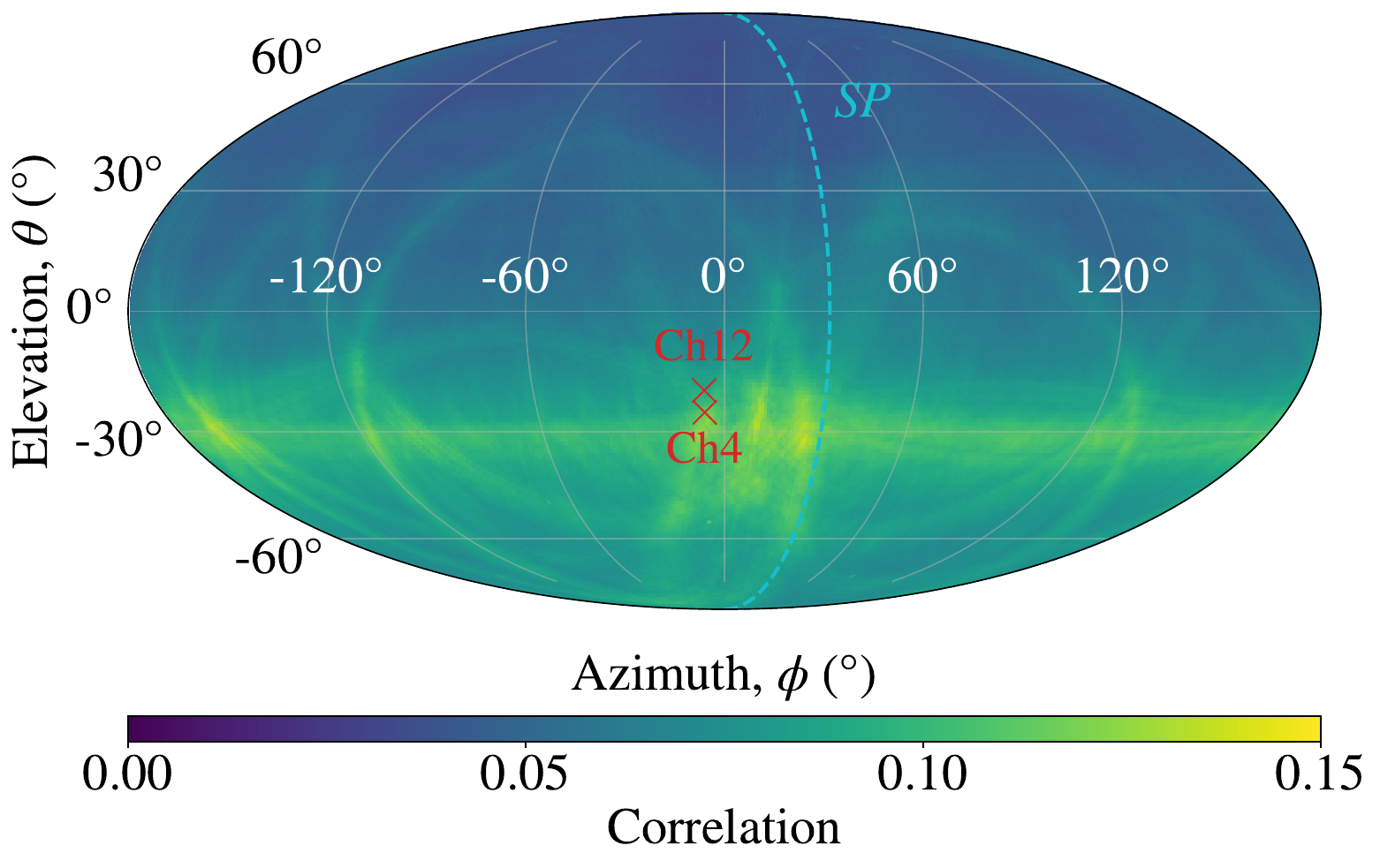}
        \centering\includegraphics[width=\linewidth]{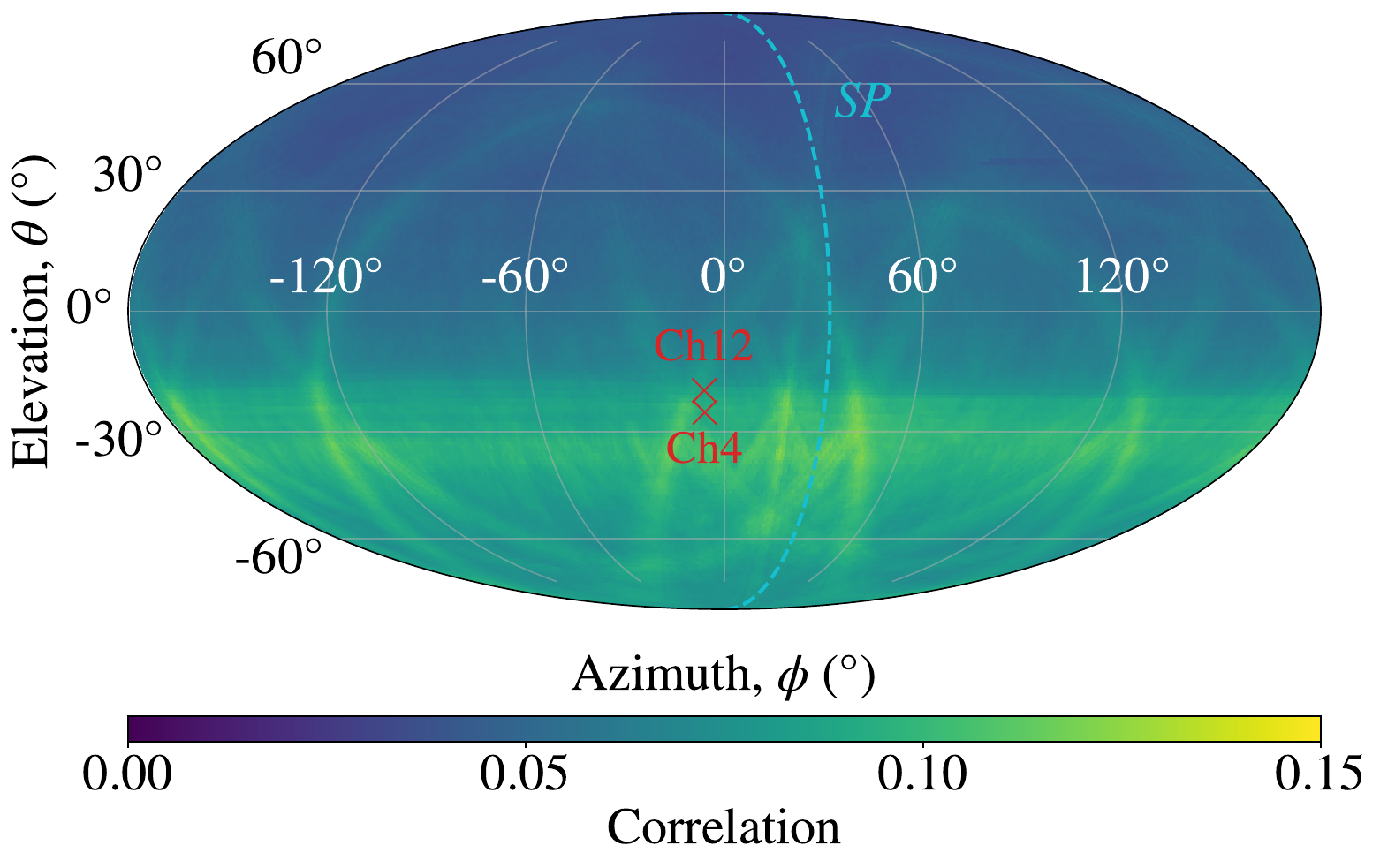}
        \caption{Joint A5-PA reconstruction map of the passing PA event assuming emission from a radius of $51.79$~m (top), corresponding to the distance to the local calibration pulser, and $300$~m (bottom), corresponding to a distant source. The direction to channels 4 and 12 on A5 (red crosses) and to South Pole (SP, dashed cyan line) are shown. The reconstruction profiles over an unknown time offset between the A5 and PA channels, choosing the offset which maximizes the correlation in each direction.}
        \label{fig:passing_event_joint_reco_alts}
\end{figure}

The total envelope is the sum of the impulsive and noise components:  $h_{\trm{tot}} = h_{\trm{impulse}} + h_{\trm{noise}}$.
Under these assumptions, $H_{\trm{tot}}(t_\trm{frac})$ can be calculated analytically from Eq.~\eqref{eq:impulsivity_cdf} as
\begin{align}
    H_{\trm{tot}}(t_\trm{frac}) = \frac{ A\,\mathrm{erf}(t_\trm{frac}/B)+t_\trm{frac} }{ A\,\mathrm{erf}(1/B)+1 }~.
\end{align}
Large $A$ and small $B$ mean the error function turns on sharply, indicating an impulsive signal.
We fit this model to each event, and calculate
the goodness-of-fit $\chi^2$, which is the summed-squared difference
between the data and the model summed across all time bins.
We fit every event's CSW twice:
once where $A=0$, to represent the pure noise null hypothesis ($\chi^2_{\trm{noise}}$), and a second time allowing $A$ and $B$ to float freely ($\chi^2_{\trm{impulse}}$) in an impulsive alternative hypothesis.
The difference between them $\Delta\chi^2 = \chi^2_{\mathrm{noise}} - \chi^2_{\mathrm{impulse}}$ is converted into a statistical significance $\xi_{\trm{tot}}$ using Wilk's theorem, given two additional degrees of freedom.
In this way, $\xi_{\trm{tot}}$ represents the significance of the preference for the alternative hypothesis (impulse + noise) over the pure noise null hypothesis.
The final three parameters provided to the LDA are $B$, $\chi^2_{\trm{impulse}}$, and $\xi_{\trm{tot}}$.

\section{Distant PA Detector Emission Event Reconstructions}
\label{app:alt_recos}

The detector emission event observed by PA, presented in Section~\ref{sec:results} and Fig.~\ref{fig:passing_event_trace}, is best reconstructed assuming a local source hypothesis. Figure~\ref{fig:passing_event_joint_reco} shows a reconstruction assuming an emission radius of $32.93$~m from the station center, corresponding to the distance to channels 4 and 12. Here we show the reconstruction maps when using alternative assumptions for the emission radius. Specifically, we present the standard reconstruction radii used in our analysis of PA data: $51.79$~m, the distance to the local calibration pulser, and $300$~m, our proxy for a distant source. These reconstructions are presented in Fig.~\ref{fig:passing_event_joint_reco_alts}. As the hypothesized emission radius increases, the maximum correlation decreases, and the quality of the reconstruction degrades: the clear peak in Fig.~\ref{fig:passing_event_joint_reco} dissolves into several local maxima at $51.79$~m before smearing into a broad band at $300$~m. This behavior provides strong evidence that the signal shown in Fig.~\ref{fig:passing_event_trace} originates from a point within the A5 station.

\bibliography{main}

@article{Ahlers:2012rz,
    author = "Ahlers, Markus and Halzen, Francis",
    title = "{Minimal Cosmogenic Neutrinos}",
    eprint = "1208.4181",
    archivePrefix = "arXiv",
    primaryClass = "astro-ph.HE",
    doi = "10.1103/PhysRevD.86.083010",
    journal = "Phys. Rev. D",
    volume = "86",
    pages = "083010",
    year = "2012"
}

@article{ANITA:2018impulsivity,
    author = "Gorham, P. W. and others",
    collaboration = "ANITA",
    title = "{Constraints on the diffuse high-energy neutrino flux from the third flight of ANITA}",
    eprint = "1803.02719",
    archivePrefix = "arXiv",
    primaryClass = "astro-ph.HE",
    doi = "10.1103/PhysRevD.98.022001",
    journal = "Phys. Rev. D",
    volume = "98",
    number = "2",
    pages = "022001",
    year = "2018"
}

@article{ANITA:2019wyx,
    author = "Gorham, P. W. and others",
    collaboration = "ANITA",
    title = "{Constraints on the ultrahigh-energy cosmic neutrino flux from the fourth flight of ANITA}",
    eprint = "1902.04005",
    archivePrefix = "arXiv",
    primaryClass = "astro-ph.HE",
    doi = "10.1103/PhysRevD.99.122001",
    journal = "Phys. Rev. D",
    volume = "99",
    number = "12",
    pages = "122001",
    year = "2019"
}

@article{ARA:2011testbed,
    author = "Allison, P. and others",
    title = "{Design and Initial Performance of the Askaryan Radio Array Prototype EeV Neutrino Detector at the South Pole}",
    eprint = "1105.2854",
    archivePrefix = "arXiv",
    primaryClass = "astro-ph.IM",
    doi = "10.1016/j.astropartphys.2011.11.010",
    journal = "Astropart. Phys.",
    volume = "35",
    pages = "457--477",
    year = "2012"
}

@article{ARA:2018phased,
    author = "Allison, P. and others",
    title = "{Design and performance of an interferometric trigger array for radio detection of high-energy neutrinos}",
    eprint = "1809.04573",
    archivePrefix = "arXiv",
    primaryClass = "astro-ph.IM",
    doi = "10.1016/j.nima.2019.01.067",
    journal = "Nucl. Instrum. Meth. A",
    volume = "930",
    pages = "112--125",
    year = "2019"
}

@article{ARA:2019wcf,
    author = "Allison, P. and others",
    collaboration = "ARA",
    title = "{Constraints on the diffuse flux of ultrahigh energy neutrinos from four years of Askaryan Radio Array data in two stations}",
    eprint = "1912.00987",
    archivePrefix = "arXiv",
    primaryClass = "astro-ph.HE",
    doi = "10.1103/PhysRevD.102.043021",
    journal = "Phys. Rev. D",
    volume = "102",
    number = "4",
    pages = "043021",
    year = "2020"
}

@article{ARA:2022rwq,
    author = "Allison, P. and others",
    collaboration = "ARA",
    title = "{Low-threshold ultrahigh-energy neutrino search with the Askaryan Radio Array}",
    eprint = "2202.07080",
    archivePrefix = "arXiv",
    primaryClass = "astro-ph.HE",
    doi = "10.1103/PhysRevD.105.122006",
    journal = "Phys. Rev. D",
    volume = "105",
    number = "12",
    pages = "122006",
    year = "2022"
}

@article{Fang:2013vla,
    author = "Fang, Ke and Kotera, Kumiko and Murase, Kohta and Olinto, Angela V.",
    title = "{Testing the Newborn Pulsar Origin of Ultrahigh Energy Cosmic Rays with EeV Neutrinos}",
    eprint = "1311.2044",
    archivePrefix = "arXiv",
    primaryClass = "astro-ph.HE",
    doi = "10.1103/PhysRevD.90.103005",
    journal = "Phys. Rev. D",
    volume = "90",
    number = "10",
    pages = "103005",
    year = "2014",
    note = "[Erratum: Phys.Rev.D 92, 129901 (2015)]"
}

@article{IceCube:2025ewu,
    author = "Abbasi, R. and others",
    collaboration = "IceCube",
    title = "{Improved measurements of the TeV-PeV extragalactic neutrino spectrum from joint analyses of IceCube tracks and cascades}",
    eprint = "2507.22234",
    archivePrefix = "arXiv",
    primaryClass = "astro-ph.HE",
    doi = "10.1103/4n6v-r7n4",
    journal = "Phys. Rev. D",
    volume = "113",
    pages = "062002",
    year = "2026"
}

@article{IceCubeCollaborationSS:2025jbi,
    author = "Abbasi, R. and others",
    collaboration = "IceCube",
    title = "{Search for Extremely-High-Energy Neutrinos and First Constraints on the Ultrahigh-Energy Cosmic-Ray Proton Fraction with IceCube}",
    eprint = "2502.01963",
    archivePrefix = "arXiv",
    primaryClass = "astro-ph.HE",
    doi = "10.1103/PhysRevLett.135.031001",
    journal = "Phys. Rev. Lett.",
    volume = "135",
    number = "3",
    pages = "031001",
    year = "2025"
}

@article{KM3NeT:2025npi,
    author = "Aiello, S. and others",
    collaboration = "KM3NeT",
    title = "{Observation of an ultra-high-energy cosmic neutrino with KM3NeT}",
    doi = "10.1038/s41586-024-08543-1",
    journal = "Nature",
    volume = "638",
    number = "8050",
    pages = "376--382",
    year = "2025",
    note = "[Erratum: Nature 640, E3 (2025)]"
}

@article{Kotera:2010yn,
    author = "Kotera, K. and Allard, D. and Olinto, A. V.",
    title = "{Cosmogenic Neutrinos: parameter space and detectability from PeV to ZeV}",
    eprint = "1009.1382",
    archivePrefix = "arXiv",
    primaryClass = "astro-ph.HE",
    doi = "10.1088/1475-7516/2010/10/013",
    journal = "JCAP",
    volume = "10",
    pages = "013",
    year = "2010"
}

@article{Muzio:2023skc,
    author = "Muzio, Marco Stein and Unger, Michael and Wissel, Stephanie",
    title = "{Prospects for joint cosmic ray and neutrino constraints on the evolution of trans-Greisen-Zatsepin-Kuzmin proton sources}",
    eprint = "2303.04170",
    archivePrefix = "arXiv",
    primaryClass = "astro-ph.HE",
    doi = "10.1103/PhysRevD.107.103030",
    journal = "Phys. Rev. D",
    volume = "107",
    number = "10",
    pages = "103030",
    year = "2023"
}

@article{PierreAuger:2023pjg,
    author = "Abdul Halim, Adila and others",
    collaboration = "Pierre Auger",
    title = "{Latest results from the searches for ultra-high-energy photons and neutrinos at the Pierre Auger Observatory}",
    doi = "10.22323/1.444.1488",
    journal = "PoS",
    volume = "ICRC2023",
    pages = "1488",
    year = "2023"
}

@article{Greisen1966,
    author = "Greisen, Kenneth",
    title = "{End to the Cosmic-Ray Spectrum?}",
    doi = "10.1103/PhysRevLett.16.748",
    journal = "Phys. Rev. Lett.",
    volume = "16",
    pages = "748--750",
    year = "1966"
}

@article{Zatsepin1966,
    author = "Zatsepin, G. T. and Kuzmin, V. A.",
    title = "{Upper limit of the spectrum of cosmic rays}",
    journal = "JETP Lett.",
    volume = "4",
    pages = "78--80",
    year = "1966"
}

@article{Askaryan:1961pfb,
    author = "Askaryan, G. A.",
    title = "{Excess negative charge of an electron-photon shower and its coherent radio emission}",
    journal = "Sov. Phys. JETP",
    volume = "14",
    pages = "441--443",
    year = "1962"
}

@article{ANITA:2006nif,
    author = "Gorham, P. W. and others",
    collaboration = "ANITA",
    title = "{Observations of the Askaryan Effect in Ice}",
    eprint = "hep-ex/0611008",
    archivePrefix = "arXiv",
    doi = "10.1103/PhysRevLett.99.171101",
    journal = "Phys. Rev. Lett.",
    volume = "99",
    pages = "171101",
    year = "2007"
}

@article{Barwick:2016mxm,
    author = "Barwick, S. W. and others",
    collaboration = "ARIANNA",
    title = "{Radio detection of air showers with the ARIANNA experiment on the Ross Ice Shelf}",
    eprint = "1612.04473",
    archivePrefix = "arXiv",
    primaryClass = "astro-ph.HE",
    doi = "10.1016/j.astropartphys.2017.01.009",
    journal = "Astropart. Phys.",
    volume = "90",
    pages = "50--68",
    year = "2017"
}

@article{RNO-G:2020rmc,
    author = "Aguilar, J. A. and others",
    collaboration = "RNO-G",
    title = "{Design and performance of the Radio Neutrino Observatory Greenland (RNO-G)}",
    eprint = "2010.12279",
    archivePrefix = "arXiv",
    primaryClass = "astro-ph.IM",
    doi = "10.1088/1748-0221/16/03/P03032",
    journal = "JINST",
    volume = "16",
    number = "03",
    pages = "P03032",
    year = "2021"
}

@article{ARA:2026wjc,
    author = "Alden, N. and others",
    collaboration = "ARA",
    title = "{Sensitivity of the Current Askaryan Radio Array to Ultra-High Energy Neutrinos}",
    eprint = "2605.04268",
    archivePrefix = "arXiv",
    primaryClass = "astro-ph.HE",
    month = "5",
    year = "2026",
    journal = "",
}

@article{IceCube-Gen2:2020qha,
    author = "Aartsen, M. G. and others",
    collaboration = "IceCube-Gen2",
    title = "{IceCube-Gen2: the window to the extreme Universe}",
    eprint = "2008.04323",
    archivePrefix = "arXiv",
    primaryClass = "astro-ph.HE",
    doi = "10.1088/1361-6471/abbd48",
    journal = "J. Phys. G",
    volume = "48",
    number = "6",
    pages = "060501",
    year = "2021"
}

@article{Klein:2005di,
    author = "Klein, J. R. and Roodman, A.",
    title = "{Blind analysis in nuclear and particle physics}",
    reportNumber = "SLAC-PUB-12051",
    doi = "10.1146/annurev.nucl.55.090704.151521",
    journal = "Ann. Rev. Nucl. Part. Sci.",
    volume = "55",
    pages = "141--163",
    year = "2005"
}

@article{Dziewonski:1981xy,
    author = "Dziewonski, A. M. and Anderson, D. L.",
    title = "{Preliminary reference earth model}",
    doi = "10.1016/0031-9201(81)90046-7",
    journal = "Phys. Earth Planet. Interiors",
    volume = "25",
    pages = "297--356",
    year = "1981"
}

@article{Connolly:2011vc,
    author = "Connolly, Amy and Thorne, Robert S. and Waters, David",
    title = "{Calculation of High Energy Neutrino-Nucleon Cross Sections and Uncertainties Using the MSTW Parton Distribution Functions and Implications for Future Experiments}",
    eprint = "1102.0691",
    archivePrefix = "arXiv",
    primaryClass = "hep-ph",
    doi = "10.1103/PhysRevD.83.113009",
    journal = "Phys. Rev. D",
    volume = "83",
    pages = "113009",
    year = "2011"
}

@article{Cummings:2023iuw,
    author = "Cummings, Austin and Krebs, Ryan and Wissel, Stephanie and Alvarez-Mu{\~n}iz, Jaime and Carvalho, Jr., Washington R. and Romero-Wolf, Andr{\'e}s and Schoorlemmer, Harm and Zas, Enrique",
    title = "{Secondary lepton production, propagation, and interactions}",
    eprint = "2311.03646",
    archivePrefix = "arXiv",
    primaryClass = "hep-ph",
    doi = "10.1103/PhysRevD.111.023012",
    journal = "Phys. Rev. D",
    volume = "111",
    number = "2",
    pages = "023012",
    year = "2025"
}

@article{ARA:2023rtx,
    author = "Seikh, Mohammad Ful Hossain and others",
    collaboration = "ARA",
    title = "{Calibration and Physics with ARA Station 1: A Unique Askaryan Radio Array Detector}",
    eprint = "2308.07292",
    archivePrefix = "arXiv",
    primaryClass = "astro-ph.IM",
    doi = "10.22323/1.444.1163",
    journal = "PoS",
    volume = "ICRC2023",
    pages = "1163",
    year = "2023"
}

@article{ARA:2023frx,
    author = "Dasgupta, Paramita and others",
    collaboration = "ARA",
    title = "{Progress Towards a Diffuse Neutrino Search in the Full Livetime of the Askaryan Radio Array}",
    eprint = "2308.12125",
    archivePrefix = "arXiv",
    primaryClass = "astro-ph.HE",
    doi = "10.22323/1.444.1226",
    journal = "PoS",
    volume = "ICRC2023",
    pages = "1226",
    year = "2023"
}

@phdthesis{Meures:2014rfw,
    author = "Meures, Thomas",
    title = "{Development of a sub-glacial array of radio antennas for the detection of the flux of GZK neutrinos}",
    doi = "10.1007/978-3-319-18756-3",
    school = "U. Brussels, Brussels U., IIHE",
    year = "2014"
}

@article{ARA:2021wmr,
    author = "Dasgupta, Paramita and others",
    collaboration = "ARA",
    title = "{The Calibration of the Geometry and Antenna Delay in Askaryan Radio Array Station 4 and 5}",
    doi = "10.22323/1.395.1086",
    journal = "PoS",
    volume = "ICRC2021",
    pages = "1086",
    year = "2021"
}

@article{Vieregg:2015baa,
    author = "Vieregg, A. G. and Bechtol, K. and Romero-Wolf, A.",
    title = "{A Technique for Detection of PeV Neutrinos Using a Phased Radio Array}",
    eprint = "1504.08006",
    archivePrefix = "arXiv",
    primaryClass = "astro-ph.IM",
    doi = "10.1088/1475-7516/2016/02/005",
    journal = "JCAP",
    volume = "02",
    pages = "005",
    year = "2016"
}

@article{ARA:2025qit,
    author = "Alden, N. and others",
    collaboration = "ARA",
    title = "{Observation of In-Ice Askaryan Radiation from High-Energy Cosmic Rays}",
    eprint = "2510.21104",
    archivePrefix = "arXiv",
    primaryClass = "astro-ph.HE",
    doi = "10.1103/xwqy-yzrk",
    journal = "Phys. Rev. Lett.",
    volume = "136",
    number = "15",
    pages = "151003",
    year = "2026"
}

@article{ARA:2025scq,
    author = "Seikh, Mohammad Ful Hossain and others",
    collaboration = "ARA",
    title = "{Techniques for Continuous Wave Identification and Filtering in the Askaryan Radio Array}",
    eprint = "2509.20735",
    archivePrefix = "arXiv",
    primaryClass = "astro-ph.IM",
    doi = "10.22323/1.501.1171",
    journal = "PoS",
    volume = "ICRC2025",
    pages = "1171",
    year = "2025"
}

@article{ARA:2014fyf,
    author = "Allison, P. and others",
    collaboration = "ARA",
    title = "{First Constraints on the Ultra-High Energy Neutrino Flux from a Prototype Station of the Askaryan Radio Array}",
    eprint = "1404.5285",
    archivePrefix = "arXiv",
    primaryClass = "astro-ph.HE",
    doi = "10.1016/j.astropartphys.2015.04.006",
    journal = "Astropart. Phys.",
    volume = "70",
    pages = "62--80",
    year = "2015"
}

@article{RICE:2001ayk,
    author = "Kravchenko, I. and others",
    collaboration = "RICE",
    title = "{Performance and simulation of the RICE detector}",
    eprint = "astro-ph/0112372",
    archivePrefix = "arXiv",
    doi = "10.1016/S0927-6505(02)00194-9",
    journal = "Astropart. Phys.",
    volume = "19",
    pages = "15--36",
    year = "2003"
}

@article{Barwick:2005zz,
    author = "Barwick, S. and Besson, D. and Gorham, P. and Saltzberg, D.",
    title = "{South Polar in situ radio-frequency ice attenuation}",
    doi = "10.3189/172756505781829467",
    journal = "J. Glaciol.",
    volume = "51",
    pages = "231--238",
    year = "2005"
}

@article{Gorski:2004by,
    author = "G{\'o}rski, K. M. and Hivon, E. and Banday, A. J. and Wandelt, B. D. and Hansen, F. K. and Reinecke, M. and Bartelmann, M.",
    title = "{HEALPix - A Framework for high resolution discretization, and fast analysis of data distributed on the sphere}",
    eprint = "astro-ph/0409513",
    archivePrefix = "arXiv",
    doi = "10.1086/427976",
    journal = "Astrophys. J.",
    volume = "622",
    pages = "759--771",
    year = "2005"
}

@article{Anker:2019rzo,
    author = "Anker, A. and others",
    title = "{A search for cosmogenic neutrinos with the ARIANNA test bed using 4.5 years of data}",
    eprint = "1909.00840",
    archivePrefix = "arXiv",
    primaryClass = "astro-ph.IM",
    doi = "10.1088/1475-7516/2020/03/053",
    journal = "JCAP",
    volume = "03",
    pages = "053",
    year = "2020"
}

@article{Feldman:1997qc,
    author = "Feldman, Gary J. and Cousins, Robert D.",
    title = "{A Unified approach to the classical statistical analysis of small signals}",
    eprint = "physics/9711021",
    archivePrefix = "arXiv",
    reportNumber = "HUTP-97-A096",
    doi = "10.1103/PhysRevD.57.3873",
    journal = "Phys. Rev. D",
    volume = "57",
    pages = "3873--3889",
    year = "1998"
}

@article{Allison:2015eky,
      author         = "Allison, P. and others",
      title          = "{Performance of two Askaryan Radio Array stations and
                        first results in the search for ultrahigh energy
                        neutrinos}",
      collaboration  = "ARA",
      journal        = "Phys. Rev.",
      volume         = "D93",
      year           = "2016",
      number         = "8",
      pages          = "082003",
      doi            = "10.1103/PhysRevD.93.082003",
      eprint         = "1507.08991",
      archivePrefix  = "arXiv",
      primaryClass   = "astro-ph.HE",
      SLACcitation   = "%%CITATION = ARXIV:1507.08991;%%"
}

@article{prl,
    author = "Alden, N. and others",
    collaboration = "ARA",
    title = "{New Constraints on the Ultrahigh Energy Neutrino Flux from the First Array-Wide Search with the Askaryan Radio Array}",
    eprint = "2611.14350",
    archivePrefix = "arXiv",
    primaryClass = "astro-ph.HE",
    month = "11",
    year = "2026",
    journal = ""
}

@article{Allison:2017jpy,
    author = "Allison, P. and others",
    title = "{Measurement of the real dielectric permittivity $\epsilon_r$ of glacial ice}",
    eprint = "1712.03301",
    archivePrefix = "arXiv",
    primaryClass = "astro-ph.IM",
    doi = "10.1016/j.astropartphys.2019.01.004",
    journal = "Astropart. Phys.",
    volume = "108",
    pages = "63--73",
    year = "2019"
}

@article{Coleman:2022abf,
    author = "Coleman, A. and others",
    title = "{Ultra high energy cosmic rays The intersection of the Cosmic and Energy Frontiers}",
    eprint = "2205.05845",
    archivePrefix = "arXiv",
    primaryClass = "astro-ph.HE",
    reportNumber = "FERMILAB-PUB-22-413-PPD",
    doi = "10.1016/j.astropartphys.2023.102819",
    journal = "Astropart. Phys.",
    volume = "149",
    pages = "102819",
    year = "2023"
}

@article{Roth:2025hqe,
    author = "Roth, Markus",
    collaboration = "Pierre Auger",
    title = "{Exploring the Ultra-High-Energy Universe: Highlights from the Pierre Auger Observatory}",
    doi = "10.22323/1.501.1402",
    journal = "PoS",
    volume = "ICRC2025",
    pages = "1402",
    year = "2025"
}

@article{Kido:2025mey,
    author = "Kido, Eiji",
    collaboration = "Telescope Array",
    title = "{Recent Progress of the Telescope Array Experiment}",
    doi = "10.22323/1.501.1410",
    journal = "PoS",
    volume = "ICRC2025",
    pages = "1410",
    year = "2025"
}

@article{Unger:2023lob,
    author = "Unger, Michael and Farrar, Glennys R.",
    title = "{The Coherent Magnetic Field of the Milky Way}",
    eprint = "2311.12120",
    archivePrefix = "arXiv",
    primaryClass = "astro-ph.GA",
    doi = "10.3847/1538-4357/ad4a54",
    journal = "Astrophys. J.",
    volume = "970",
    number = "1",
    pages = "95",
    year = "2024"
}

@article{PierreAuger:2024hlp,
    author = "Abdul Halim, A. and others",
    collaboration = "Pierre Auger",
    title = "{Impact of the magnetic horizon on the interpretation of the Pierre Auger Observatory spectrum and composition data}",
    eprint = "2404.03533",
    archivePrefix = "arXiv",
    primaryClass = "astro-ph.HE",
    reportNumber = "FERMILAB-PUB-24-0144-CSAID-PPD-TD-V",
    doi = "10.1088/1475-7516/2024/07/094",
    journal = "JCAP",
    volume = "07",
    pages = "094",
    year = "2024"
}

@article{Bister:2024ocm,
    author = "Bister, Teresa and Farrar, Glennys R. and Unger, Michael",
    title = "{The Large-scale Anisotropy and Flux (de)magnification of Ultrahigh-energy Cosmic Rays in the Galactic Magnetic Field}",
    eprint = "2408.00614",
    archivePrefix = "arXiv",
    primaryClass = "astro-ph.HE",
    doi = "10.3847/2041-8213/ad856f",
    journal = "Astrophys. J.",
    volume = "975",
    number = "1",
    pages = "L21",
    year = "2024"
}

@article{Roe:1998zi,
    author = "Roe, Byron P. and Woodroofe, Michael B.",
    title = "{Improved probability method for estimating signal in the presence of background}",
    eprint = "physics/9812036",
    archivePrefix = "arXiv",
    doi = "10.1103/PhysRevD.60.053009",
    journal = "Phys. Rev. D",
    volume = "60",
    pages = "053009",
    year = "1999"
}

@article{Kim:2023svg,
    author = "Kim, Myoungchul",
    collaboration = "ARA",
    title = "{Enhanced Ultra-High Energy Neutrino Search at the Askaryan Radio Array using Template-based Techniques}",
    doi = "10.22323/1.444.1148",
    journal = "PoS",
    volume = "ICRC2023",
    pages = "1148",
    year = "2023"
}

@phdthesis{hong2014,
    author = {Hong, Eugene S.},
    title  = {Searching for Ultra-high Energy Neutrinos with Data from a Prototype Station of the {Askaryan Radio Array}},
    school = {The Ohio State University},
    year   = {2014},
    url    = {http://rave.ohiolink.edu/etdc/view?acc_num=osu1417624919}
}

@article{Weigel:2024gzh,
    author = "Weigel, Philip L. R. and Conrad, Janet M. and Garcia-Soto, Alfonso",
    title = "{Cross sections and inelasticity distributions of high-energy neutrino deep inelastic scattering}",
    eprint = "2408.05866",
    archivePrefix = "arXiv",
    primaryClass = "hep-ph",
    doi = "10.1103/PhysRevD.111.043044",
    journal = "Phys. Rev. D",
    volume = "111",
    number = "4",
    pages = "043044",
    year = "2025"
}

@article{Cooper-Sarkar:2011jtt,
    author = "Cooper-Sarkar, Amanda and Mertsch, Philipp and Sarkar, Subir",
    title = "{The high energy neutrino cross-section in the Standard Model and its uncertainty}",
    eprint = "1106.3723",
    archivePrefix = "arXiv",
    primaryClass = "hep-ph",
    doi = "10.1007/JHEP08(2011)042",
    journal = "JHEP",
    volume = "08",
    pages = "042",
    year = "2011"
}

@phdthesis{clark2019,
  author = {Clark, Brian A.},
  title  = {Optimization of a Search for Ultra-High Energy Neutrinos in Four Years of Data of {ARA} Station 2},
  school = {The Ohio State University},
  address = {Columbus, OH, USA},
  year   = {2019},
  type   = {Ph.D. thesis}
}

@phdthesis{lu2020,
  author = {Lu, Ming-Yuan},
  title  = {Search for Ultra-High Energy Neutrinos with the {Askaryan Radio Array}},
  school = {University of Wisconsin--Madison},
  address = {Madison, WI, USA},
  year   = {2020},
  type   = {Ph.D. thesis}
}

@article{Allison:2019rgg,
    author = "Allison, P. and others",
    title = "{Long-baseline horizontal radio-frequency transmission through polar ice}",
    eprint = "1908.10689",
    archivePrefix = "arXiv",
    primaryClass = "astro-ph.IM",
    doi = "10.1088/1475-7516/2020/12/009",
    journal = "JCAP",
    volume = "12",
    pages = "009",
    year = "2020"
}

@phdthesis{Hughes:2022thesis,
  author    = {Hughes, Kaeli Autumn},
  title     = {A Low Threshold Neutrino Search with the {Askaryan Radio Array}},
  school    = {University of Chicago},
  year      = {2022},
  month     = jun,
  doi       = {10.6082/uchicago.4043},
  url       = {https://doi.org/10.6082/uchicago.4043}
}

@article{PUEO:2020bnn,
    author = "Abarr, Q. and others",
    collaboration = "PUEO",
    title = "{The Payload for Ultrahigh Energy Observations (PUEO): a white paper}",
    eprint = "2010.02892",
    archivePrefix = "arXiv",
    primaryClass = "astro-ph.IM",
    doi = "10.1088/1748-0221/16/08/P08035",
    journal = "JINST",
    volume = "16",
    number = "08",
    pages = "P08035",
    year = "2021"
}

@article{PUEO:2026owv,
    author = "Abarr, Q. and others",
    collaboration = "PUEO",
    title = "{The design and flight of the Payload for Ultrahigh Energy Observations (PUEO)}",
    eprint = "2610.35675",
    archivePrefix = "arXiv",
    primaryClass = "astro-ph.HE",
    month = "10",
    year = "2026",
    journal = ""
}

\end{document}